\documentclass[11pt,letterpaper,usenames,dvipsnames]{article}
\usepackage{jheppub}

\usepackage{enumitem}

\RequirePackage{pdf14}
\usepackage[utf8]{inputenc}
\usepackage{graphicx,tikz}
\usepackage[framemethod=TikZ]{mdframed} 
\usepackage{amsfonts,amsmath,amssymb,amsthm,esint}
\usepackage{array,setspace,mathrsfs,yfonts,dsfont,bbm,colonequals,amscd,mathtools}
\usepackage{relsize,suffix,cancel,bbm,capt-of,upgreek,verbatim,xcolor}
\usepackage[T1]{fontenc}
\usepackage{enumitem}
\usepackage{tcolorbox}
\usepackage{arydshln}
\usepackage{amsthm}

\newcounter{ex}[section]
\renewcommand{\theex}{\arabic{section}}
\newenvironment{ex}[2][]{%
\refstepcounter{ex}%
\ifstrempty{#1}%
{\mdfsetup{%
frametitle={%
\tikz[baseline=(current bounding box.east),outer sep=0pt]
\node[anchor=east,rectangle,fill=blue!20]
{\strut Exercises Section~\theex};}}
}%
{\mdfsetup{%
frametitle={%
\tikz[baseline=(current bounding box.east),outer sep=0pt]
\node[anchor=east,rectangle,fill=blue!20]
{\strut Exercises Section~\theex:~#1};}}%
}%
\mdfsetup{innertopmargin=10pt,linecolor=blue!20,%
linewidth=2pt,topline=true,%
frametitleaboveskip=\dimexpr-\ht\strutbox\relax
}
\begin{mdframed}[]\relax%
\label{#2}}{\end{mdframed}}

\newcounter{example}[section]
\renewcommand{\theexample}{\arabic{section}.\arabic{example}}

\newcounter{dig}[section]
\renewcommand{\thedig}{\arabic{section}.\arabic{dig}}
\newenvironment{dig}[2][]{%
\refstepcounter{dig}%
\ifstrempty{#1}%
{\mdfsetup{%
frametitle={%
\tikz[baseline=(current bounding box.east),outer sep=0pt]
\node[anchor=east,rectangle,fill=WildStrawberry!40]
{\strut Digression~\thedig};}}
}%
{\mdfsetup{%
frametitle={%
\tikz[baseline=(current bounding box.east),outer sep=0pt]
\node[anchor=east,rectangle,fill=WildStrawberry!40]
{\strut Digression~\thedig:~#1};}}%
}%
\mdfsetup{innertopmargin=10pt,linecolor=WildStrawberry!40,%
linewidth=2pt,topline=true,%
frametitleaboveskip=\dimexpr-\ht\strutbox\relax
}
\begin{mdframed}[]\relax%
\label{#2}}{\end{mdframed}}

\newcounter{definition}[section]
\renewcommand{\thedefinition}{\arabic{section}.\arabic{definition}}
\newenvironment{definition}[2][]{%
\refstepcounter{definition}%
\ifstrempty{#1}%
{\mdfsetup{%
frametitle={%
\tikz[baseline=(current bounding box.east),outer sep=0pt]
\node[anchor=east,rectangle,fill=NavyBlue!40]
{\strut Definition~\thedefinition};}}
}%
{\mdfsetup{%
frametitle={%
\tikz[baseline=(current bounding box.east),outer sep=0pt]
\node[anchor=east,rectangle,fill=NavyBlue!40]
{\strut Definition~\thedefinition:~#1};}}%
}%
\mdfsetup{innertopmargin=10pt,linecolor=NavyBlue!40,%
linewidth=2pt,topline=true,%
frametitleaboveskip=\dimexpr-\ht\strutbox\relax
}
\begin{mdframed}[]\relax%
\label{#2}}{\end{mdframed}}

\newcounter{solution}[section]
\renewcommand{\thesolution}{\arabic{section}.\arabic{solution}}
\newenvironment{solution}[2][]{%
\refstepcounter{solution}%
\ifstrempty{#1}%
{\mdfsetup{%
frametitle={%
\tikz[baseline=(current bounding box.east),outer sep=0pt]
\node[anchor=east,rectangle,fill=ForestGreen!40]
{\strut Solution~\thesolution};}}
}%
{\mdfsetup{%
frametitle={%
\tikz[baseline=(current bounding box.east),outer sep=0pt]
\node[anchor=east,rectangle,fill=ForestGreen!40]
{\strut Solution~\thesolution:~#1};}}%
}%
\mdfsetup{innertopmargin=10pt,linecolor=ForestGreen!40,%
linewidth=2pt,topline=true,%
frametitleaboveskip=\dimexpr-\ht\strutbox\relax
}
\begin{mdframed}[]\relax%
\label{#2}}{\end{mdframed}}

\def\dblue#1{{\color{NavyBlue}{#1}}}

\def\dgreen#1{{\color{ForestGreen}{#1}}}

\def\red#1{{\color{red}{#1}}}
\def\dred#1{{\color{OrangeRed}{#1}}}

\def\bsa{{\boldsymbol a}}

\def\bt{{\boldsymbol \t}}

\def\tq{{\til q}}

\def\ub{{\bar u}}

\def\red#1{{\color{red}{#1}}}
\def\del{{\partial}}
\def\delb{{\bar\del}}
\def\bar{\overline}
\def\til{\widetilde}

\def\vev#1{{\langle{#1}\rangle}} 

\def\^{\wedge}
\def\I{\mathds{1}}
\def\Tr{\mathop{\rm Tr}}
\def\Im{\mathop{\rm Im}}

\def\U{{\rm U}}

\def\SO{\mathop{\rm so}}
\def\SL{\mathop{\rm SL}}
 
\def\Sp{\mathop{\rm sp}}

\def\C{\mathbb{C}} 
\def\tCs{\til{\C^*}}

\def\N{\mathbb{N}}

\def\R{\mathbb{R}} 
 
\def\Z{\mathbb{Z}}

\def\mH{\mathsf{H}}

\def\a{{\alpha}}
\def\ad{{\dot\a}}

\def\b{{\beta}}

\def\g{{\gamma}}

\def\d{{\delta}}

\def\D{{\Delta}}
\def\e{{\epsilon}}

\def\z{{\zeta}}
\def\th{{\theta}}
\def\thd{{\theta^\dag}}

\def\l{{\lambda}}

\def\L{{\Lambda}}
\def\m{{\mu}}

\def\s{{\sigma}}
\def\sb{{\bar{\sigma}}}
\def\S{{\Sigma}}
\def\t{{\tau}}

\def\bPhi{{\boldsymbol \Phi}}

\def\w{{\omega}}

\def\ff{{\mathfrak f}}
\def\gf{{\mathfrak g}}

\def\hf{{\mathfrak h}}
\def\iif{{\mathfrak i}}
\def\jf{{\mathfrak j}}
\def\Mf{{\mathfrak M}}

\def\sof{\mathfrak{so}}
\def\spf{\mathfrak{sp}}
\def\suf{\mathfrak{su}}

\def\cB{{\mathcal B}}

\def\cC{{\mathcal C}}
\def\cCrg{{\mathcal C}_{\rm reg}}

\def\cE{{\mathcal E}}
\def\cF{{\mathcal F}}
\def\cG{{\mathcal G}}
\def\cH{{\mathcal H}}
\def\cI{{\mathcal I}}

\def\cL{{\mathcal L}}
\def\cM{{\mathcal M}}
\def\cN{{\mathcal N}}

\def\cO{{\mathcal O}}

\def\cQ{{\mathcal Q}}

\def\cT{{\mathcal T}}
\def\cTg{{\mathcal T}[\gf]}
\def\cV{{\mathcal V}}
\def\cW{{\mathcal W}}

\def\Csc{\mathscr{C}}
\def\Msc{\mathscr{M}}

\def\beq{\begin{equation}}
\def\eeq{\end{equation}}
\def\bea{\begin{eqnarray}}
\def\eea{\end{eqnarray}}

\def\dt{{\rm d}^2\theta}
\def\dtd{{\rm d}^2\theta^\dag}
\def\dtb{{\rm d}^2\theta^\dag}

\def\tcH{\tilde{\cH}}
\def\shm{\text{-}}

\def\Ssw{\S_{\rm SW}}
\def\lsw{\l_{\rm SW}}

\def\sCB{\phi}

\newcommand{\bpmat}{\begin{pmatrix}}
\newcommand{\epmat}{\end{pmatrix}}
\newcommand{\bsmat}{\begin{smallmatrix}}
\newcommand{\esmat}{\end{smallmatrix}}

\newtheorem{exe}{Exercise}

\title{The constraining power of Coulomb Branch Geometry: \emph{lectures on Seiberg-Witten theory}}
\author[1]{Mario Martone}
\affiliation[1]{University of Texas, Austin, Physics Department, Austin TX 78712}
\emailAdd{mariomartone@utexas.edu}

\abstract{The constraining mathematical structure of the Coulomb branch of four dimensional $\cN=2$ supersymmetric theories is  discussed. The presentation follows a somewhat different route from other excellent reviews on the subject and it is geared towards using this tool to classify four dimensional $\cN=2$ superconformal field theories. This is the writeup of the lectures given at the Winter School ``YRISW 2020'' to appear in a special issue of JPhysA.
}

\begin{document}
\maketitle


\section*{Introduction}

Over the last decades, it has become increasingly clearer that the space of Quantum Field Theories is vastly larger than initially thought. These developments have been driven by a dramatic improvement in our understanding of quantum field theory in the strongly coupled regime using both numerical and analytic techniques. These lectures will review a tiny part of the latter and only within the context of four dimensional supersymmetric field theories.

Non-lagrangian field theories, that is quantum field theories which are not naturally presented via their lagrangian description, and for which such a description is often not currently known, are at the heart of the large increase of the number of known supersymmetric field theories.\footnote{With this definition, any strongly-coupled fixed point is an example of a non-Lagrangian field theory.} For instance, if we further constrain our problem and restrict to superconformal field theories where we can make educated statements about the relative size of field theories with a given property, we have the following situation as a function of the amount of (global) supersymmetry:\footnote{Of course the considerations below only reflect the perspective of the author.} 
\begin{itemize}

\item[$\cN=1$] The space of $\cN=1$ superconformal field theories is very rich but not constrained enough. A complete classification seems unlikely to be completed soon (see \cite{Razamat:2020pra} and reference therein, for some very interesting recent developments).

\item[$\cN=2$] Very rich as well as constrained. A complete classification is hard but achievable. The space of $\cN=2$ superconformal field theories is populated overwhelmingly by non-largangian field theories and the existence of a lagrangian description appears instead to be the exception \cite{Argyres:1995xn,Argyres:1995jj,Gaiotto:2009we,Gaiotto:2009hg}. These lectures will delve in some of techniques which can be used to obtain these results.

\item[$\cN=3$] Recently discovered \cite{Garcia-Etxebarria:2015wns,Aharony:2015oyb,Argyres:2016xua,Aharony:2016kai}. Great progress has already been made \cite{Bonetti:2018fqz,Argyres:2019ngz,Argyres:2019yyb}, and a complete classification might be close, though awaits the development of string theory techniques to construct $\cN=3$ theories systematically. All $\cN=3$ theories are non-lagrangian and superconformal.

\item[$\cN=4$] Theories with maximal supersymmetry are very constrained, it is believed that they are completely classified by gauge Lie algebras $\gf$ and the spectrum of electric and magnetic line operators \cite{Aharony:2013hda}, plus possible discrete gauging \cite{Bourget:2018ond,Argyres:2018zay}.  All $\cN=4$ theories are then lagrangian and superconformal.

\end{itemize}

The reader that is not already familiar with a notion of a non-lagrangian field theory, might wonder how, if not via its lagrangian, can a field theory be defined. There are many ways to answer this question but we will focus on one. The presence of supersymmetry generically allows for ground state configurations parametrized by a set of continuous variables which can in turn be interpreted as coordinates of a space called the \emph{moduli space of vacua} ($\cM$). $\cM$ inherits a very rigid mathematical structure from supersymmetry (and other symmetry in the theory) and many physical properties of the theory itself can be extracted directly from $\cM$. The situation is so constrained that, in many cases, the existence of a given field theory can be purely inferred by the consistency of $\cM$ without a need for a description in terms of weakly coupled fields.

The focus of these lectures is to discuss the structure of a particular subset of $\cM$; the Coulomb Branch of four dimensional $\cN=2$ supersymmetric field theories. The understanding of this space was revolutionized by the two seminal papers by Seiberg and Witten \cite{Seiberg:1994rs,Seiberg:1994aj} which grew in what has become known as Seiberg-Witten theory. I will also refer to the totality of the geometric constraints on the Coulomb branch captured by the Seiberg-Witten analysis as \emph{Special Kahler geometry}. There is by now an incredibly vast literature on the subject and it is therefore impossible to be systematic in appropriately reference so many brilliant contributions. I apologize in advance for the omissions that I might have made. 

More specifically, in these notes I will try to give a somewhat pedagogical introduction to the main ideas of Seiberg-Witten theory. My presentation will be motivated by the classification spirit outlined above and I will therefore present the topic in a slightly unconventional way. For example I will \emph{not} discuss the infamous monopole-dyon singularities appearing on the Coulomb branch of $\cN=2$ $\suf(2)$ super Yang-Mills theory but rather focus on discussing the Coulomb branch of IR-free or conformal theories. 
For a more standard treatment of the subject the reader can consult the many wonderful existing reviews of the subject, e.g. \cite{AlvarezGaume:1996mv,Lerche:1996xu,Argyres:1996nonp}.

The lectures are organized as follows. The first lecture introduces the moduli space of vacua of $\cN=2$ gauge theories $\cM$ and focuses in particular on its CB $\cC$. I will also discuss the simplest examples in some details. The second lecture focuses instead on introducing the special coordinates, the K\"ahler metric and $\cC$ as a K\"ahler space. In the third lecture we will analyze the abstract geometric structure introduced in the first two lectures in a concrete example. By explicit calculation, we will familiarize with the puzzling fact that the effective low-energy theory on $\cC$ does not have a globally defined lagrangian description and, relatedly, the CB has metric singularities. In the fourth lecture, we will introduce the concept of electromagnetic duality which resolves in a non-trivial fashion the puzzle of a non-global lagrangian description on $\cC$. In the fifth lecture we will introduce formally the notion of Special K\"ahler geometry which describes the geometric structure of the CB. We then use this framework to derive all the consistent scale invariant one complex dimensional CB geometries. Through this exercise we will argue for the existence of many $\cN=2$ superconformal field theories which do not have a lagrangian formulation. I will conclude in lecture six by introducing the Seiberg-Witten (SW) curve and one-form, as well as outlining the generalization to higher ranks and providing a brief description on how the CB geometry is constructed in the large zoo of $\cN=2$ theories which can be obtained from compactification of 6$d$ (2,0) theories, the so called class-S theories.

Throughout these lecture notes we made various digressions which lie somewhat off the main line of arguments of the manuscript and could therefore be skipped at a first read. We chose to leave them there as they might appeal to the more advanced readers. We have also assumed knowledge of $\cN=1$ supersymmetry which is not reviewed here. There are by now plenty of fantastic references e.g.\cite{Ramond:1999susy,Weinberg:2000III,Terning:2006ms,Dine:2007Sst,Shifman:2012lec,Gates:1983nr,Shifman:1995ua,Martin:1997ns,Argyres:2001eva,Luty:2005sn,Aitchison:2005cf}

Finally, these lectures are vastly incomplete. The list of subjects we had to omit, given the time constraints of the school and of life in general, is so long that we can't even be complete in listing our incompletness. It is possible that future versions of this manuscript might partially integrate these omissions. In particular with the addition a discussion of massive BPS states and mass deformations, as well as one on the relations of Coulomb branch geometry across dimensions, in particular drawing connection between the four dimensional story and the one in six, five and three dimensions.

\section{Moduli space of vacua of theories}\label{sec:modsp}
\setcounter{exe}{0}

We will start by introducing the main protagonist of these lectures: \emph{the Coulomb branch of the moduli space of vacua} of four dimensional $\cN=2$ supersymmetric field theories. Supersymmetric field theories generally admit continuous vacuum configurations. These vacua can be interpreted as coordinates of a space which we call ``the moduli space of vacua''. Depending on the amount of supersymmetry, these spaces are increasingly more constrained. We will restrict here to the case of $\cN=2$ supersymmetry and later we will also add the extra constrain of superconformal invariance. Henceforth we will use the letter $\cM$ to refer to the entire moduli space but, as we will describe momentarily, $\cM$ decomposes in different branches with different geometric properties, see figure \ref{MS}. In this first lecture we will familiarize with the general structure and introduce these different branches which take the name of \emph{Coulomb}, \emph{Higgs} and \emph{Mixed} branch. 

\begin{figure}
\begin{center}
\includegraphics[width=.65\textwidth]{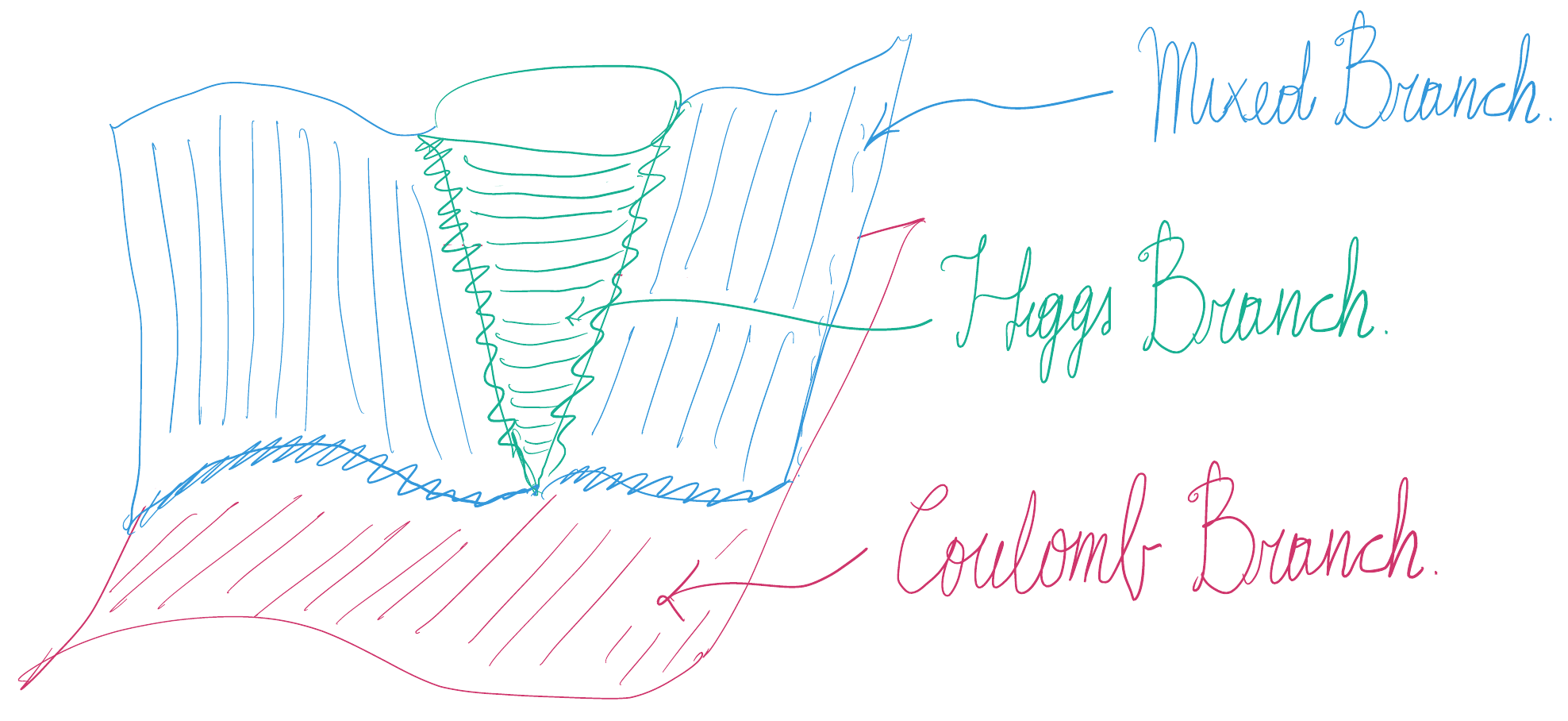}
\caption{Moduli space of vacua of a four dimensional $\cN=2$ supersymmetric theory.}
\label{MS}
\end{center}
\end{figure}

\subsection{$\cN=2$ Lagrangian}

We here remind the basics of $\cN=2$ supersymmetric field theories and establish nomenclature and some conventions. For more systematic reviews see e.g. \cite{Argyres:2001eva,Tachikawa:2013kta}. The $\cN=2$ super-Poincar\'e algebra takes the schematic form:
\beq\label{N2Susy}
\{Q^n_\alpha,Q^\dag_{\dot{\alpha}m}\}=2\delta_n^m\sigma^\mu_{\alpha\dot{\alpha}} P_{\mu},\qquad \{Q^n_{\a},Q^m_{\b}\}=\epsilon_{\a\b}\epsilon_{nm}Z\qquad [Q^n_\a,P_{\mu}]=0.
\eeq
The $Q$ are are left-handed Weyl spinors and the $P_\mu$ are the four momentum which generates translation. The commutator between the $Q$s and the rest of the generators of the Poincar\'e group follow from their representations. We use the following conventions for the matrices $\sigma^\mu$ and $\sb^\mu$ are Pauli matrices for $\mu=1,2,3$ and the identity for $\mu=0$:
\beq
\sigma^0=\sb^0=\left(
\begin{array}{cc}
1&0\\
0&1
\end{array}
\right),\quad
\sigma^1=\shm\sb^1=\left(
\begin{array}{cc}
0&1\\
1&0
\end{array}
\right),\quad
\sigma^2=\shm\sb^2=\left(
\begin{array}{cc}
0&-i\\
i&0
\end{array}
\right),\quad
\sigma^3=\shm\sb^3=\left(
\begin{array}{cc}
1&0\\
0&-1
\end{array}
\right).
\eeq 

\eqref{N2Susy} has a non-trivial $\suf(2)_R\times \U(1)_R$ R-symmetry group. To discuss the vacuum structure of $\cN=2$ supersymmetric theories we need to first introduce the lagrangian of a $\cN=2$ gauge theory with gauge Lie algebra $\gf$. The Lagrangian is written down in terms of two $\cN=2$ supermultiplets the $\cN=2$ \emph{vector multiplet} and the $\cN=2$ \emph{hypermultiplet}:
\begin{subequations}
\begin{align}
\cN=2\ \text{Vector multiplet}\quad&:\quad\left\{
\begin{array}{c}
\sCB^A\\
\l^A_\a\qquad \tilde\l^A_\a\\
A^A_\mu
\end{array}
\right.,\\
\cN=2\ \text{Hypermultiplet}\quad&:\quad\left\{
\begin{array}{c}
\psi^i_\a\\
q^i\qquad \tilde q^i\\
\tilde \psi_i^\ad
\end{array}
\right.,
\end{align}
\end{subequations}
where fields on the same row fall in representations of $\suf(2)_R$ and we wrote explicitly the spinor indices. The fields of the vector and hyper multiplet also falls in $\cN=1$ representations, which are arranged diagonally above. The $\cN=2$ vector multiplet can be decomposed as an $\cN=1$ vector and chiral multiplets: $V^{\cN=2}=(\Phi,V^{\cN=1})$. The hypermultiplet can instead be decomposed in a chiral and anti-chiral $\cN=1$ super-multiplets: $\mH=(\cH_i,\tcH^i)$. Since the $\cN=2$ vector multiplet transforms in the adjoint representation, we also wrote explicitly the gauge index $A=1,...,{\rm dim}\gf$. The lower-case index $i=1,...,N_{\mH}$ is instead a flavor index.


Imposing the invariance under super-gauge transformations:
\beq
\left.
\begin{array}{c}\vspace{.3em}
\Phi\to\left(e^{i\Omega} \right)|_{\bf R} \Phi,\\\vspace{.3em}
\cH_i\to\left(e^{i\Omega} \right)|_{\bf R} \cH_i,\\\vspace{.3em}
\tcH_i\to\left(e^{i\Omega} \right)|_{\bf \bar R} \tcH_i,\\\vspace{.3em}
e^V\to e^{i\Omega^\dag}e^Ve^{-i\Omega}.
\end{array}
\right\}\qquad
\Omega=2 T^A\Omega_A
\eeq
where $\Omega$ is a $\cN=1$ chiral superfield and $T^A\in \gf$, the Lagrangian follows:
\beq\label{N2L}
\cL=\frac{1}{4\pi}{\rm Im}\left[\int\dt\dtb {\rm tr}\left(\Phi^\dag e^V \Phi+\cH_i^\dag e^V \cH^i+\tcH^{\dag i} e^V\tcH_i\right)+\tau\int\dt\left(\frac{1}{2}{\rm tr} W^2+\cW\right)\right]
\eeq
The explicit expression of \eqref{N2L} in terms of individual bosonic and fermionic degrees of freedom, is derived in exercise \ref{Lcomp}. If you haven't done it at least once in your life, we strongly advice you to work out this form in detail. It is a pedantic yet instructive exercise. In the expression for the lagrangian, $\tau$ is the \emph{holomorphic $\gf$ gauge coupling} :
\beq
\tau=i\frac{4\pi}{g^2}+\frac{\theta}{2\pi}.
\eeq
and the closely resembling $W$ and $\cW$ should not be confused. The first one is the standard chiral superfield obtained out of the $\cN=1$ vector superfield $V$
\beq\label{ChiVec}
W_\alpha:=-\frac14\overline{DD}D_\a(V)
\eeq 
while the second one is the $\cN=1$ superpotential which is fixed by $\cN=2$ supersymmetry to be
\beq\label{SupTer}
\cW=\sqrt{2}\tilde{\cH}_i\Phi \cH^i+\sqrt{2} m^i_j\tilde{\cH}_i\cH^j.
\eeq

The first term in the lagrangian is given by the integral of the entire $\cN=1$ superspace and is what we will call in the following the \emph{Kahler potential}:
\beq\label{KaPotN2}
K=\frac{1}{4\pi}{\rm Im}\left[\int\dt\dtb {\rm tr}\left(\Phi^\dag e^V \Phi+\cH_i^\dag e^V \cH^i+\tcH^{\dag i} e^V\tcH_i\right)\right].
\eeq
Now that we are done with the basics, can start our real analysis. The first question we need to answer is what are the classical vacua of this theory. This is easily addressed by looking at the zeros of the scalar potential which looks:
\beq\label{ScaPot}
V=\frac{1}{2}D^AD_A+F^{\iif\dag}F_{\iif},\qquad {\rm with} \qquad \left\{\def\arraystretch{1.3}
\begin{array}{l}
D^A=\varphi^\dag_{\iif}T^{A\iif}_\jf\varphi^\jf\\
F_\iif=\frac{\partial\cW}{\partial \varphi^\iif}
\end{array}
\right.
\eeq
where the $\iif$ runs over a range such that $\varphi^\iif$ label all the dynamical scalar fields of the theory. Here $A=1,...,{\rm dim} \gf$ indicate the elements of the Lie algebra of $\gf$. From exercise \ref{FDterm} we derive that the equations to find the vacua of the theory are:
\beq\label{eqmot}
\def\arraystretch{1.2}
\begin{array}{c|c}
\textsc{F-term}&\textsc{D-term}\\
\hline
\hline
\quad (q^i\tq_i)_{\rm traceless}=0\quad{}&[\sCB,\sCB^\dag]=0\\
q^im_i^j+\sCB q^j=0&\quad[q^iq^\dag_i-\tq_i\tq^{\dag i}]_{\rm traceless}=0\quad{}\\
m_j^i \tq_i+\tq^j \sCB=0&
\end{array}
\eeq
To simplify the discussion in the following we will set $m^i_j=0$. In this limit there are two general ``types'' of solutions to the equations \eqref{eqmot}:

\begin{enumerate}[leftmargin=3.8cm,rightmargin=.3cm]
\itemsep0em 
\item[\textbf{\textsc{\dred{Coulomb Branch}}}] $q=\tq=0$, $\sCB\neq0$.  The name Coulomb branch (CB) derives from the fact that low-energy limit description of an $\cN=2$ supersymmetric theory on its CB is $\U(1)^r$ $\cN=2$ gauge theory, where $r$ is the rank of the theory. This branch of the moduli space will be the main focus of these lectures and we will indicate it as $\cC\subset\cM$.

\item[\textbf{\textsc{\dgreen{Higgs Branch}}}] $\sCB=0$ and $q\neq\tq\neq0$. Generally the gauge group is fully Higgsed along this branch, thus the name Higgs branch. The low-energy effective descriprion of an $\cN=2$ theory is simply given by free decoupled hypermultiplets. We won't discuss this branch in any real detail. 
\end{enumerate}

\begin{enumerate}[leftmargin=3.8cm,rightmargin=.3cm]

\item[\textbf{\textsc{\dblue{Mixed Branch}}}] $\sCB\neq q\neq\tq\neq0$. Finding the general solution of the equations in \ref{eqmot} is quite challenging. They play an important role in connecting the constraints from \emph{Higgs} and \emph{Coulomb} branches.
\end{enumerate}

\noindent A depiction of the moduli space of vacua of a general four dimensional $\cN=2$ gauge theory is provided in fig. \ref{MS}. 

\subsection{Rank-1: $\suf(2)$}

Let's start familiarizing with the structure of the CB in the simplest example, an $\cN=2$ gauge theory with $\gf=\suf(2)$. 

Setting $q=\tq=0$, a solution for the scalar of the vector multiplet which satisfies \eqref{eqmot} can be explicitly written as:
\beq\label{CBSol}
\sCB_{\rm CB}=\left(
\begin{array}{cc}
a&0\\
0&-a
\end{array}
\right)
\eeq
This vev breaks $\suf(2)\to U(1)$ implying that the effective theory at the generic point of the CB is a $\cN=2$ $U(1)$ theory. $\suf(2)$ is rank 1 therefore this matches the general claim made above. In the presence of hypermultiplets, for $\sCB_{\rm CB}\neq0$, the superpotential term in \eqref{SupTer} gives a mass to all the hypers:
\beq
m_{\rm CB}\sim |a|
\eeq  
therefore at low enough energies, the hypermultiplets also decouple and the theory on a generic point of the CB is a $\cN=2$ $U(1)$ theory with \emph{no charged matter}. This theory is not interacting and it is described by a single $\cN=2$ vector multiplet. So far the situation seems quite boring, yet, perhaps surprisingly, the detailed description of this (extremely elementary) low-energy abelian theory gives instead a surprisingly rich geometric structure to $\cC$ and will be the focus of a large part of the following lectures.

The form \ref{CBSol} does not fix the gauge redundancy completely and the remaining gauge transformations impose further identifications on $\sCB_{\rm CB}$ which we have to account for. From exercise \ref{SolCB}, the extra gauge redundancy in the case of $\suf(2)$ reduces to a $\Z_2$; $\sCB_{\rm CB}$ and $\shm\sCB_{\rm CB}$ are gauge equivalent. Since $a$ is not gauge invariant, to properly parametrize this branch of the moduli space, we need to define instead the following gauge invariant quantity:
\beq\label{uCB}
u={\rm tr}\ \sCB_{\rm CB}^2=2 a^2
\eeq
$u$ will be our gauge invariant coordinate on the CB which, for $\suf(2)$, is a one complex dimensional. Henceforth we will drop the subscript $_{\rm CB}$ when referring to the vev of the scalar component of the vector superfield on the CB. It will hopefully be clear from the context when we refer to the scalar field and when to its vev.

The HB of $\suf(2)$ theories depends on the number of the hypermultiplets and its description is considerably trickier. For completion and to satisfy the curiosity of the interested reader, we report a brief discussion of the HB of the $\cN=2$ $\suf(2)$ theory with four hypermultiplets in the fundamental below. This theory is superconformal and thus its description is somewhat easier and gives a taste on how to use a more abstract approach, strongly relying on the algebra of gauge invariant operators, to discuss moduli spaces of $\cN=2$ theories. This discussion lies somewhat outside the main theme of the lectures and can be safely skipped during your first read.


\begin{dig}[Higgs branch geometry of $\cN=2$ $\suf(2)$ with $N_f=4$]{SU2}

Since the fundamental representation of $\suf(2)$ is pseudo-real, it follows that the $\cN=2$ $\suf(2)$ gauge theory with four hypermultiplets in the fundamentals has an $\SO(8)$ flavor symmetry, enlarged from the naive $U(4)$ guess. Explicitly, the Pauli matrices satisfy
\beq
(\sigma^{i})^*=-\sigma_2\sigma^i\sigma_2
\eeq
which allow the re-arrangement of $(q^i,\tq_i)$, $i=1,...,4$ into a fundamental of $\SO(8)$. We will label it as $\bar{q}_\ell$ such that $\bar{q}$ contains both $q$ and $\tq$, appropriately arranged to form an ${\bf 8}$ of $\SO(8)$. It is handy to describe the HB directly in terms of the gauge invariant, meson, operator which can be obtained from $\bar{q}_\ell$:
\beq\label{MesOp}
\Mf^{\ell_1}_{\ell_2}:=\bar{q}^{\dag\ell_1}_a\bar{q}_{\ell_2}^a
\eeq
where we made explicit the $\suf(2)$ gauge contraction which, recall, are done via the antisymmetric $\epsilon_{ab}$ symbol. $\Mf$ is therefore anti-symmetric in $\ell_1$ and $\ell_2$ and it transforms in the adjoint representation of the $\SO(8)$ or the ${\bf 28}$. This implies that $\Mf^2:= \Mf^{\ell_1}_{\ell_2}\Mf^{\ell_3}_{\ell_4}$ tranform in the Sym$_{\otimes^2}{\bf 28}$ and which can in turn be decomposed in terms of a direct sum of $\SO(8)$ irreducible representations:
\beq\label{repsM}
{\rm Sym}_{\otimes^2}{\bf 28}={\bf 300}\oplus {\bf 35}_v\oplus {\bf 35}_c\oplus {\bf 35}_s\oplus {\bf 1}
\eeq

The fact that the ``quark'' fields $\bar{q}$ satisfy \eqref{eqmot} implies that $\Mf$ satisfies non-trivial relations \cite{Argyres:1996eh} and which can most easily formulated in terms of $\SO(8)$ representation theory. The idea is that some of the IRREPs appearing by decomposing monomials of the $\Mf$s vanish. It turns out that for this particular case, all the relation appear at the quadratic order and can be framed in terms of vanishing representation in \eqref{repsM}. Specifically the F-term conditions are:
\beq
\Mf\otimes\Mf|_{{\bf 35}_v}=0\qquad\Mf\otimes\Mf|_{\bf 1}=0
\eeq
Furthermore from the fact that we write $\Mf$ in terms of the quark fields, it follows that $\Mf$ automatically satisfies:
\beq
\Mf\otimes\Mf|_{{\bf 35}_s}=0\qquad\Mf\otimes\Mf|_{{\bf 35}_c}=0
\eeq
We then conclude that the Higgs branch of this theory can be fully recast in the following quadratic relation for the meson field:
\beq
\Mf^2|_{\bf 300}\neq0
\eeq

\end{dig}

\subsection{Rank-2: $\suf(3)$}

Now let us turn to the analysis of the second simplest example: $\suf(3)$ $\cN=2$ gauge theories. Again, since we are interested in general statements about $\cC$, we will not specify the hypermultiplet content and what we will say apply broadly. In this case the CB solution has the form:
\beq\label{CBSolSU3}
\sCB_{\suf(3)}=\left(
\begin{array}{ccc}
a_1&0&0\\
0&a_2&0\\
0&0&-a_1-a_2
\end{array}
\right)
\eeq
from which it immediately follows that in this case the CB is a two complex dimensional space.

If $a_1\neq a_2$, $\suf(3)\to U(1)\times U(1)$ again confirming the general statement above. As before, the configuration in \eqref{CBSolSU3} \emph{is not} gauge invariant. In this case, figuring out the set of identifications imposed on \eqref{CBSolSU3} by the remaining gauge redundancy is more involved but with enough effort it can be shown that is tantamount of shuffling the eigenvalues around. This group is nothing but the symmetric group $S_3$. To identify the correct, globally defined, two complex coordinate on $\cC$, we need to consider polynomial in $a_1$ and $a_2$ which are invariant under this $S_3$ which are just the symmetric polynomials in $a_1$ and $a_2$ 

A possible convenient basis of the symmetric polynomials in $a_1$ and $a_2$ is readily obtained considering ${\rm tr}\ \sCB_{\suf(3)}^2$ and ${\rm tr}\ \sCB_{\suf(3)}^3$:
\beq
u_1:=\frac{{\rm tr}\ \sCB_{\suf(3)}^2}{2}=a_1^2+a_2^2+a_1a_2,\qquad u_2:=\frac{{\rm tr}\ \sCB_{\suf(3)}^3}{3}=-a_1^2 a_2-a_1a_2^2
\eeq
$(u_1,u_2)$, which have scaling dimension respectively 2 and 3, are the global coordinates on $\cC$ generalizing \eqref{uCB}.  Here the description of the low energy theory is a bit trickier. It remains true that on a generic point of $\cC$ all the hypers acquire a mass and thus the low-energy theory is a $U(1)^2$ with no charged degrees of freedom, this time described by two $\cN=2$ vector multiplet. But in this case, there are non-trivial direction in the $(a_1,a_2)$ space for which the masses of the hypermultiplets vanish. The details depend on the superpotential \eqref{SupTer} which is fixed by the number of hypermultiplets and their representations. We will ignore this subtlety for now and only take home the (partially incorrect) message that the low energy theory is $\cN=2$ theory of free photons.

\begin{dig}[Mixed branch of $\suf(3)$ with $N_f=6$]{SU3}
$\cN=2$ $\suf(3)$ theories are the theories that provide the simplest non-trivial example of a mixed branch which we will discuss below for the case with $N_f=6$ which is in fact superconformal. Again this discussion is largely inessential to understand the rest of these lectures and can be skipped.  

Since the fundamental representation of $\suf(3)$ is a complex representation, this theory has an $\suf(6)$ flavor symmetry which acts non-trivially on the hypermultiplets. The HB structure is more involved than the $\suf(2)$ case and we will not discuss it, for more details see for example \cite{Gaiotto:2008nz}. Interestingly this theory has a non-trivial \emph{Mixed Branch} (MB) which is relatively easy to analyze and we will discuss it here. Again it is convenient to describe this branch in terms of a gauge invariant operator: $\widetilde \Mf$:
\beq\label{MesSU3}
\widetilde \Mf_i^{\phantom{i}j}:=\tq_i^aq^j_a
\eeq

For generic value of $a_1$ and $a_2$, the F-term conditions set the vev of the squarks $(q,\tq)$, and therefore that of $\widetilde \Mf$, to zero. When any of the entries in \eqref{CBSolSU3} vanish, the squarks can acquire a non zero vev ``along'' that same direction, with the remaining two components vanishing. Since both $q$ and $\tq$ are forced by the F-terms to have their non vanishing vev ``aligned'', we can treat them as they are simple complex numbers and the F-term conditions further simplify. Then the structure of the mixed branch of the $\suf(3)$ case can again be encoded in a set of quadratic relations satisfied by the $\widetilde \Mf$. Explicitly $\widetilde \Mf^2$ transforms
\beq
{\rm Sym}_{\otimes^2}{\bf 35}={\bf 405}\oplus {\bf 189}\oplus {\bf 35}\oplus {\bf 1}
\eeq
imposing the F-term conditions and the fact that $\widetilde \Mf$ is written in terms of the quarks as in \eqref{MesSU3}, implies that 
\beq
\widetilde \Mf^2|_{{\bf 189}\oplus{\bf 35}\oplus {\bf1}}=0\qquad{\rm and}\qquad \widetilde \Mf^2|_{{\bf 405}}\neq0
\eeq
It can be checked, though it is not at all obvious by the equation above, that these relations identify a ten complex-dimensional variety. 

Even though there are three different choices of $a_1$ and $a_2$ which allows for the diagonal entry to vanish, the three are physically equivalent as they are related by Weyl transformations. Thus the invariant way of describing the mixed branch as a subvariety of the two-dimensional CB is as the one complex dimensional plane $u_2=0$. Along this subvariety the effective low-energy theory is a $U(1)\times U(1)$ and the mixed branch can be identified as the Higgs branch of one of the two $U(1)$ factor for which some of the quarks charged under it can acquire a vev.

\end{dig}

\subsection{CB in the general case}

Let's conclude this section describing the CB of a $\cN=2$ $\suf(n)$ gauge theory. In this case \eqref{CBSol} and \eqref{CBSolSU3} generalizes to:
\beq
\phi=\left(
\begin{array}{ccccc}
a_1&0&0&...&0\\
0&a_2&0&...&0\\
\vdots&&\ddots&&0\\
0&&...&&-\sum_i a_i
\end{array}
\right),\qquad a_1,...,a_{n-1} \in \C
\eeq
and for $a_1\neq a_2\neq...\neq a_{n-1}$ $\suf(n)\to \U(1)^{n-1}$ and the CB is an $n-1$ complex dimensional space. As before the $a_i$ are not gauge invariant and the pattern of discrete identifications which we have seen in our previous examples generalizes straightforwardly to $\suf(n)$. The remaining gauge redundancy acts shuffling the eigenvalues around and therefore closes on the symmetric group $S_n$. The gauge invariant coordinates on $\cC$ are:
\beq
u_1={\rm Tr}\big[\phi^2\big],\quad u_2={\rm Tr}\big[\phi^3\big], \quad \cdots,\quad u_{n-1}={\rm Tr}\big[\phi^n\big].
\eeq 
which have scaling dimension $\D_{u_i}=i-1$.

\begin{table}[t]
\centering
$\begin{array}{c|cc}
\gf & \ \text{Weyl}(\gf)\ & \ |\text{Weyl}(\gf)| \ \\
\hline
\suf(r+1)\ {\rm or}\ A_r & S_{r+1} & \ (r+1)!  \ \\
\spf(2r)/\sof(2r+1)\ {\rm or}\ BC_r & \ S_r \ltimes \Z_2^r \ & 2^r r! \\
\sof(2r)\ {\rm or}\ D_r & \ S_r \ltimes \Z_2^{r-1} \ & 2^{r-1} r! \\
G_2 & \ \Z_2 \ltimes S_3 \ & 2^2\cdot 3 \\
F_4 & \ S_3 \ltimes \text{Weyl}(D_4)\ & 2^7 \cdot 3^2 \\
E_6 & \ldots & \ 2^7\cdot 3^4\cdot 5 \ \\ 
E_7 & \ldots & \ 2^{10}\cdot 3^4\cdot 5 \cdot 7 \ \\
E_8 & \ldots & 2^{14}\cdot 3^5 \cdot 5^2 \cdot 7  
\end{array}$ 
\caption{\label{weylinfo} The Weyl group for each Lie algebra $\gf$.}
\end{table}

This structure can easily generalized to any rank-$r$ gauge Lie algebra $\gf$. Setting the hypermultiplet scalar to zero reduces the equation of motion to
\beq\label{solgenCB}
[\sCB^\dag,\sCB]=0\quad
\eeq
which implies that the CB vacuum configuration are parametrized by the complex scalar of the vector multiplet restricted to the Cartan subalgebra $\hf$ of $\gf$. Calling $H_i$ a choice of the generators of $\hf$:
\beq
 \sCB=\tilde{\sCB}_iH^i,\quad i=1,...,r\quad{\rm where}\quad [H^i,H^j]=0
\eeq
Since the $H^i$ all commute with each other they can be simultaneously diagonalized, as we did in \eqref{CBSol} and \eqref{CBSolSU3}. We will call the $r$ eigenvalues of $\sCB$, following the same notation as before, $a_1,...,a_{r}$. 

As now know, gauge redundancy introduces further identification on \eqref{solgenCB}. This set of identifications has very sharp interpretation in Lie algebra theory and coincides with the Weyl group of the corresponding Lie algebra (thus Weyl$\big(\suf(2)\big)\equiv \Z_2$, Weyl$\big(\suf(3)\big)\equiv S_3$ and Weyl$\big(\suf(n)\big)\equiv S_n$). These groups are well known for all Lie algebras \cite{mckay1981tables}, see table \ref{weylinfo}. The CB for a $\cN=2$ theory with gauge lie algebra $\gf$ is therefore parametrized by invariant Weyl$(\gf)$ polynomials in the $a_i$s and a good set of global coordinates on $\cC$ can be chosen to be the set of the generators of these polynomials. For a Lie algebra $\gf$, there are precisely $r$ generators.

The two main take away messages from this section are the following general properties of $\cN=2$ CBs:\\

\begin{tcolorbox}
\begin{center}
\textbf{Coulomb branch take aways}
\end{center}\vspace{-.7em}
\begin{itemize}[leftmargin=*]
\item[\textbf{\dred{1.}}] The CB is a $r$ complex dimensional variety parametrized by $a_i$'s, the eigenvalues of \eqref{solgenCB}. Generalizing this fact, we define the \emph{\textbf{rank}} of $\cN=2$ (superconformal) theory as the \emph{\textbf{complex dimensionality}} of its CB. To date all known rank zero $\cN=2$ are free.

\item[\textbf{\dred{2.}}] On a generic point of the CB $\gf\to U(1)^{r}$.

\end{itemize}
\end{tcolorbox}\vspace{0.5em}

\subsection{Global vs. local}\label{locglo}

At this stage, it would be very natural for the reader to see the $a_i$s and $u_i$s on the same footage and basically interchangeably.  Most of these lectures are geared towards convincing you that they are not and try to sharpen the distinction between these two objects. For now it is important to think of the gauge invariant coordinate $u_i$ as the proper coordinate on $\cC$ and the vev of the scalar component of $\sCB_{\rm CB}$, the $\cN=2$ vector multiplet on $\cC$, as instead functions of the $u_i$. When we do that, we will immediately notice some important differences among the $a_i$s and the $u_i$s

For example in the $\suf(2)$ case, we saw that:
\beq
a=\sqrt{u/2}
\eeq
The careful reader should immediately notice that $a$ \emph{is not} single-valued in $u$. As we loop around $u=0$ on $\cC$, $a$ does not come back to itself. This shows that $a$ is only a good \emph{local} coordinate on $\cC$ while $u$ does not suffer of the same pathology and is in fact \emph{globally} defined. In this example dragging $a$ along a loop only makes it change its sign, which as we discussed means that it picks a non-trivial gauge transformation. For the $\suf(3)$, and more generally $\gf$, this feature generalizes straightforwardly with the result that generically the $a_i$s, when dragged along a closed loop $\g$, don't return to their original value but to one which is gauge equivalent:
\beq
\left(
\begin{array}{c}
a_1\\
\vdots\\
a_r
\end{array}
\right)\xrightarrow{\rm along \ \g} {\rm Weyl}(\gf)\circ 
\left(
\begin{array}{c}
a_1\\
\vdots\\
a_r
\end{array}
\right)
\eeq 
where the appropriate action of the Weyl group on the $a_i$s, indicated by $\circ$, is set by the action of the gauge group on the complex scalar $\phi$. 

In the following we will see that in general the $a_i$s can suffer even more dramatic transformations along closed loops. But to build to this result we need to first analyze more systematically the low-energy limit of the an $\cN=2$ theory on its CB.
\begin{ex}[]{}
\begin{exe}\label{Lcomp}
Expand \eqref{N2L} in components and show that
\bea\label{compo}
{\cal L}\,=&-&\frac{1}{4g^2}F^a_{\mu\nu}F^{a\mu\nu}+\frac{\theta} 
{32\pi^2}F^a_{\mu\nu}\widetilde F^{a\mu\nu}
-\frac{i}{g^2} \,\lambda^a \sigma^\mu D_\mu \bar{\lambda}^a 
+\frac{1}{2g^2}D^a D^a  \nonumber\\
&+& (\partial_\mu \phi-iA^a_\mu T^a \phi)^\dagger(\partial^\mu \phi-iA^{a\mu} 
T^a \phi) -i\,\bar{\psi} 
\bar{\sigma}^\mu (\partial_\mu\psi -iA^a_\mu T^a \psi)\nonumber\\
&-& D^a \phi^\dagger T^a \phi -i\sqrt{2}\,A^\dagger T^a \lambda^a
\psi+i\sqrt{2}\, \bar{\psi} T^a \phi \bar{\lambda}^a + F^\dagger_i F_i
\nonumber\\ 
&+&\frac{\partial {\cal W}}{\partial \phi_i} \, F_{i} +
\frac{\partial \bar{\cal W}}{\partial \phi^\dagger_i} \, F^\dagger_{i} -
\frac{1}{2}\,\frac{\partial^2 {\cal W}}{\partial \phi_i\partial \phi_j}\,
\psi_{i}\psi_{j}-\frac{1}{2}\,\frac{\partial^2 \bar{\cal W}} 
{\partial \phi_i^\dagger\partial
  \phi_j^\dagger}\,\bar{\psi}_{i}\bar{\psi}_{j}\,. 
\eea
\end{exe}

\begin{center}
\rule[1mm]{2cm}{.4pt}\hspace{1cm}$\circ$\hspace{1cm} \rule[1mm]{2cm}{.4pt}
\end{center}

\begin{exe}\label{FDterm}
Using the explicit form for the $\cN=2$ lagrangian \eqref{compo}, and \eqref{SupTer}, derive that the equations which constrain the allowed vacuum configurations for an $\cN=2$ gauge theory are: 
\begin{align}
m_j^i \tq_i+\tq^j \sCB&=0&[\sCB,\sCB^\dag]&=0\\
q^im_i^j+\sCB q^j&=0&(q^iq^\dag_i-\tq_i\tq^{\dag i}&)_{\rm traceless}=0\\
(q^i\tq_i)_{\rm traceless}&=0&
\end{align}
where $\sCB$, $q$ and $\tq$ are, respectively, the scalar fields of the $\cN=2$ vector multiplet, the chiral and anti-chiral of the hypermultiplet.
\end{exe}

\begin{center}
\rule[1mm]{2cm}{.4pt}\hspace{1cm}$\circ$\hspace{1cm} \rule[1mm]{2cm}{.4pt}
\end{center}

\begin{exe}\label{SolCB}
Show that for any $\sCB$ satisfying $[\sCB^\dag,\sCB]=0$ there always exist a $h\in\suf(2)$ such that
\beq
\sCB_{\rm CB}=h \sCB h^{-1}=\left(
\begin{array}{cc}
a&0\\
0&-a
\end{array}
\right)
\eeq
Also find the one-parameter family of $\suf(2)$ transformations, $\tilde{h}(\th)$, implementing the transformation $\sCB_{\rm CB}\to-\sCB_{\rm CB}$.
\end{exe}
\end{ex}


\section{CB IR Effective theory}
\setcounter{exe}{0}


Let us look at the effective theory on $\cC$ more closely. Consider again a rank $r$ $\cN=2$ supersymmetric gauge theory with gauge algebra $\gf$ and let us analyze what happens when $\gf$ is spontaneously broken by a non-zero $\langle\sCB\rangle$. Far from the points where two eigenvalues of $\langle\sCB\rangle$ coincide, the only massless fields are the vector supermultiplets associated with the unbroken subgroup $U(1)^{r}$ of $\gf$. Breaking $\gf\to U(1)^r$ gives mass to the vector multiplets associated to the charged vector bosons and the superpotential term \eqref{SupTer} generically gives a mass to the hyper multiplets, if they are present. The massless fields don't carry any charge under the unbroken $U(1)^{r}$ and this theory will be a theory of $r$ non-interacting $\cN=2$ vector multiplets. We will see below that this seemingly trivial low-energy physics defines in fact a rich geometric structure on $\cC$.

\subsection{Prepotential}

The most general effective action (the part with at most two derivatives) of a $\cN=2$ super Yang-Mills theory is fully determined by an object called the $\cN=2$ \emph{prepotential} ${\cal F}$ which is only a function of $r$ massless vector
supermultiplets.  In the $\cN=1$ language, the corresponding Lagrangian
takes the form (with $\Phi_i$ denoting the chiral superfield component of the $\cN=2$ vector multiplets):
\bea
{\cal L}&=&\frac{1}{8\pi}{\rm Im}\left(\int d^2\theta\, {\cal F}_{ab}(\Phi) 
W^{a\alpha}W^b_\alpha + 2\int d^2\theta d^2\thd\, (\Phi^\dagger 
e^{2gV})^a {\cal F}_a(\Phi)\right)\,. \label{N=2}
\label{N=2general}
\eea
Here, ${\cal F}_a(\Phi)=\partial{\cal F}/\partial\Phi^a$,
${\cal F}_{ab}(\Phi)=\partial^2{\cal F}/\partial\Phi^a\partial\Phi^b$. From the
above, and comparing with \eqref{N2L} and \eqref{KaPotN2}, we can easily read off the K\"{a}hler potential as Im$(\Phi^{\dag a}\cF_a(\Phi))$. As we will discuss in more details below, this gives rise to a metric 
$g_{ab}={\rm Im}(\partial_a\partial_b{\cal F})$ on the space of fields. 
A metric of this form is called a special K\"{a}hler metric.
If we demand renormalisability, then ${\cal F}$ has to be quadratic in
$\Phi$, however, if we want to write a
low-energy effective action, renormalisability is not a criterion
and ${\cal F}$ can have a more complicated form. In particular, we can
start from a microscopic theory corresponding to a
quadratic prepotential, and try to construct the modified ${\cal F}$
for the low-energy Wilsonian effective action. The exact determination
of this function is the subject of the work of Seiberg and Witten \cite{Seiberg:1994aj,Seiberg:1994pq} and, partially, these lectures.

We can use this formalism to describe the low-energy limit of any $\cN=2$ supersymmetric theory on its CB. Indeed this effective action will be given by the most generic low-energy lagrangian describing a set of $r$ decoupled $\cN=2$ vector multiplet associated to the massless photons of the unbroken $\U(1)^r$ gauge theory. Specifying \eqref{N=2general} to the case of $\U(1)^r$ we get:
\beq\label{Leff}
\cL^{U(1)^{r}}_{eff}=\frac{1}{4\pi} \, {\rm Im} \, \left [ \int \, d^{4}\theta \,
\frac{\partial {\cal F}}{\partial \Phi^{i}} \bar{\Phi}^{i} +
\int \, d^{2} \theta \, \frac{1}{2}
\frac{\partial^{2}{\cal F}}{\partial \Phi^{i} \partial \Phi^{j}}
W^{\alpha i} W^j_\alpha
\right ]\,.
\eeq
In this abelian case, the Kahler potential takes the explicit form:
\beq\label{KaPot}
K(\Phi_i,\Phi_i^*)={\rm Im}\left(\bar{\Phi}_{i}\partial{\cal F}(\Phi_i)/\partial \Phi_{i}\right).
\eeq 

We have therefore reduced the problem of understanding the detail of the low-energy effective description of an $\cN=2$ supersymmetric field theory to determining the prepotential $\cF$. The way we will be able to fix $\cF$ will be extremely non-trivial and along the way we will unveil a rich geometric structure which captures the CB geometry in its full glory!

\subsection{Special Coordinates}

The effective low-energy theory \eqref{Leff} on a generic point of the CB can, in principle, be obtained by integrating out all the massive
modes as well as massless modes above a low-energy cutoff. In
practice, this procedure is not easy to implement. Seiberg and Witten \cite{Seiberg:1994rs,Seiberg:1994aj}
realized that the global structure of the CB geometry provides an indirect procedure to study the effective theory on the CB and often it can be determined exactly. 

To start, let's consider a rank-1 example where the following patter of spontaneous symmetry breaking takes place $\suf(2)\to U(1)$. The effective low-energy Lagrangian only depends on a single chiral superfield $\Phi$ and from the prepotential $\cF$ we can compute the value of the holomorphic gauge coupling associated to the low-energy $U(1)$ theory on a generic point of the CB. Recalling \eqref{CBSol}:
\beq
\tau (a) = \frac{\partial^{2} {\cF}}{\partial a^{2}}.
\label{tau}
\eeq
This generalizes to a rank $r$ theory:
\beq
\boldsymbol{\tau}({\boldsymbol a}):=\tau_{IJ} (a_I) = \frac{\partial^{2} {\cF}}{\partial a_I\partial a_J}.
\eeq

It is now time to introduce one of the key concepts of this course; the \emph{special coordinates}. \eqref{Leff} can be re-written introducing the dual of $\Phi$:
\beq\label{LeffD}
\cL_{eff}=\frac{1}{4\pi} \, {\rm Im} \, \left [ \int \, d^{4}\theta \,
\Phi^D_i\bar{\Phi}^{i} + \int \, d^{2} \theta \, \frac{1}{2} \frac{\partial\Phi^D_i}{\partial \Phi^j} W^{\alpha i} W^j_\alpha\right]\qquad \Phi_i^D:=\frac{\partial \cF}{\partial\Phi^i}
\eeq
Furthermore the scalar component of $\Phi^D$ can also acquire a vev which we will label as $a^D$. For rank-1 theories we will write this as:
\beq\label{adual}
a^D:=\frac{\partial \cF}{\partial a}\quad \Rightarrow\quad \tau(a)=\frac{\partial a^D}{\partial a^{\phantom D}}
\eeq
the pair $(a^D,a)$ is defined to be the \emph{special coordinates} on $\cC$. We have now defined three different objects ($u,a,a^D)$ that are all related to parametrizing the CB. Since the only global holomorphic coordinate on $\cC$ is $u$, $a$ and $a^D$ should be seen as holomorphic functions of $u$. 

At tree level the prepotential is quadratic in $\Phi$ and $a^D$ is classically linear in $a$. In a $\cN=2$ gauge theory the pre-potential only receives perturbative contributions at one-loop. This translates readily into the one-loop expression for $a^D$ which allows us to compute $a^D$ explicitly, at least in the regime where non-perturbative contributions can be safely neglected. It is very helpful to compute the expression in a concrete example and we will do it in the next section.

For a rank $r$ theory, $\cC$ is an $r$ complex dimensional space parametrized by $\{u_i\}$, $i=1,...,r$ and on a generic point of $\cC$ we can naturally define a set of $2r$ holomorphic special coordinates:
\beq
\s:=\left(
\begin{array}{c}
{\boldsymbol a}^D\\
{\boldsymbol a}
\end{array}
\right)
\eeq
This definition might not appear very well motivated at this point. Nor the word \emph{dual}. Soon enough we will (hopefully) address this concern.

\subsection{K\"ahler metric}



To abstract a step further, when analyzing the theory on the moduli space $\cM$, it is helpful to think of the various complex scalars of our theory as maps from space-time to this (target) space $\cM$. Restricting the K\"ahler potential \eqref{KaPotN2} to the scalar components of the chiral fields $\Phi_i$, call this restriction $K(\phi_i,\phi_i^*)$ \eqref{KaPotN2} defines a real scalar function on $\cM$ which we can use to define a metric on $\cM$:
\beq\label{metModSp}
g_{m\bar{m}}:=\partial_m\partial_{\bar{m}}K(\sCB,\sCB^*)
\eeq
where $\partial_m=\partial/\partial \sCB^m$ and $\partial_{\bar{m}}=\partial/\partial \sCB^{*m}$ and it gives rise to the line elements is $ds^2=g_{m\bar{m}}d\sCB^m\otimes d\sCB^{*\bar{m}}$. A field redefinition which preserves the chiral nature of the fields
\beq\label{phitr}
\phi^m\to f^m(\phi)
\eeq 
acts as a complex coordinate transformation on $\cM$ and gives $\cM$ naturally the structure of a (complex) manifold. It is possible to check that under \eqref{phitr}, \eqref{metModSp} transforms consistently as a metric tensor which is furthermore Kahler. $\cM$ is therefore naturally described as \emph{K\"alher manifold}.

Going back to our simple $\suf(2)$ example the K\"ahler potential acquires the particularly simple form \eqref{KaPot} and the metric \eqref{metModSp} can be readily computed to be:
\beq
ds^2 = g_{i\bar{j}}\,da^i\otimes d\bar{a}^j =
{\rm Im}\,\frac{\partial^2{\cal F}}{\partial a \partial a}
\, da \otimes d\bar{a} \,.
\eeq
Where, recall, the $a$ is the eigenvalue of the general solution \eqref{CBSol} for the scalar field of $\cN=2$ vector multiplet on $\cC$. 

This straightforwardly generalizes to a theory of rank-$r$. In this case the metric on the CB has the general form:
\beq\label{metrRr}
ds^2 = g_{i\bar{j}}\,da^i\otimes d\bar{a}^j =
{\rm Im}\,\frac{\partial^2{\cal F}}{\partial a_i \partial a_j}
\, da^i \otimes d\bar{a}^j \quad i=1,...,r
\eeq
In both cases, the components of the metric have a particularly simple and useful form. Written in this basis $g_{i\bar j}$ are the same as the generalized holomorphic gauge coupling 
\beq\label{metric}
g_{i\bar j}=\tau_{ij}(a)=\frac{\partial^2{\cal F}}{\partial a_i \partial a_j}=\frac{\partial a^D_i}{\partial a^{j\phantom{D}}} .
\eeq
Of course this relationship between $g_{i\bar j}$ is not basis independent. If we perform a coordinate transformation we will in general spoil this beautiful geometric interpretation of the low-energy $\U(1)^r$ couplings. This is at least one reason which makes the $a$s special and holding up to its name. This relationship is not modified in the quantum theory.   


\section{An example in detail}\label{sec3}
\setcounter{exe}{0}

It is helpful to work out the details of the CB geometry in a concrete example by explicitly integrating out degrees of freedom and computing the low-energy description of the $U(1)$ theory on $\cC$ from first principles. To this end we need to remind the reader about RG-running.

\subsection{RG-running and coupling in the IR} Because of quantum corrections, the gauge coupling $g$ changes with the energy scale. At one-loop the renormalization group equation is:
\beq\label{running}
\mu\frac{d g}{d\mu}=-b\frac{g^3}{16\pi^2}+\cO(g^5)
\eeq
where $\mu$ is the energy scale of the physics we are interested in understanding and $b$ is the one-loop beta function coefficient for the gauge coupling:
\beq
b=\frac{11}{6}T({\rm adj})-\frac13\sum_fT({\bf R}_f)-\frac16\sum_bT({\bf R}_b)
\eeq
where the first sum runs over the Weyl fermion in the theory and the second on the complex bosons. The contributions at higher order in $g$,  $\cO(g^5)$, correspond to higher loops. 

Neglecting higher loop corrections, the solution of \eqref{running} can be found straightforwardly:
\beq\label{StrScale}
\frac{1}{g^2(\mu)}=\frac{b}{8\pi^2}\log\left(\frac{\mu}{\Lambda}\right), \qquad \Lambda\equiv\mu_0 e^{-8\pi/bg^2(\mu_0)}
\eeq
where $\Lambda$ is called the \emph{strong coupling} or \emph{transmutation} scale and has the important physical significance of being the scale at which physics becomes strongly coupled. In \eqref{StrScale}, $\mu_0$ is any fixed scale for which $g^2(\mu_0)$ is known.

Depending on the sign of coefficient of the beta function $b$, we can identify three different behaviors:\\

\begin{tcolorbox}
\vspace{-0.5cm}
\begin{align}\nonumber
b<0\ :&\ \emph{IR-free}&&\left\{
\begin{array}{l}
\small\text{The gauge coupling runs to zero at low-energy}\\
\small\text{while it diverges at some energy scale $\L_{\rm UV}$},\\
\small\text{for this reasons these theories are not\ }\emph{UV complete}.\\
\end{array}\right.
\\\nonumber
b>0\ :&\ \emph{Asymptotically free}&&\left\{
\begin{array}{l}
\small\text{The gauge coupling diverges at some energy $\L_{\rm IR}$ but it can}\\
\small\text{instead be defined at arbitrary high energy. Perturbative}\\
\small\text{methods are not effective to investigate it low-energy limit}.\\
\end{array}\right.\\\nonumber
b=0\ :&\  \emph{Conformal}&&\left\{
\begin{array}{l}
\small\text{The gauge coupling does not run. The vanishing of the}\\
\small\text{coefficient of the beta function is usually not stable}\\
\small\text{under quantum corrections.}\\
\end{array}\right.
\end{align}
\end{tcolorbox}\vspace{0.5em}

In an $\cN=2$ theory, the holomorphic gauge coupling only receives perturbative contributions at one-loop. So $\t$ will have the general form:
\beq\label{oneloop}
\tau(\tilde{\Lambda},\mu)=-\frac{b}{2\pi i}\log\left(\frac{\mu}{\tilde{\Lambda}}\right)+\sum_{n=1}^\infty a_n \left(\frac{\tilde{\Lambda}}{\mu}\right)^{b n}
\eeq
where the non-logarithmic part comes from non-perturbative contributions which are in general hard to compute. How can we use this expression to compute the low-energy coupling on a generic point of $\cC$?

Let's first fix $\gf$ to be $\suf(2)$ and consider the case in which the theory is IR free. Recall the expression for the coefficient of the beta function \eqref{N2bfun}:
\beq
b=4-\sum_hT({\bf R}_h)
\eeq
where we used the fact that, in the normalization in which $T({\bf 2})=1$, $T({\bf 3}\equiv{\rm Adj})=4$. We then readily deduce that if we add more than five hypermultiplets in the ${\bf 2}$ then $b<0$ and our theory flows at infinite weak coupling in the IR\footnote{Of course IR-free theories can also be obtained by adding any number of hypermultiplets in any representations larger than the ${\bf 3}$.}. Notice that non-perturbative corrections in \eqref{oneloop} can be neglected so long as we consider the theory at energy scales $\mu\ll|\tilde{\Lambda}|$.

\begin{figure}
\begin{center}
\includegraphics[width=.6\textwidth]{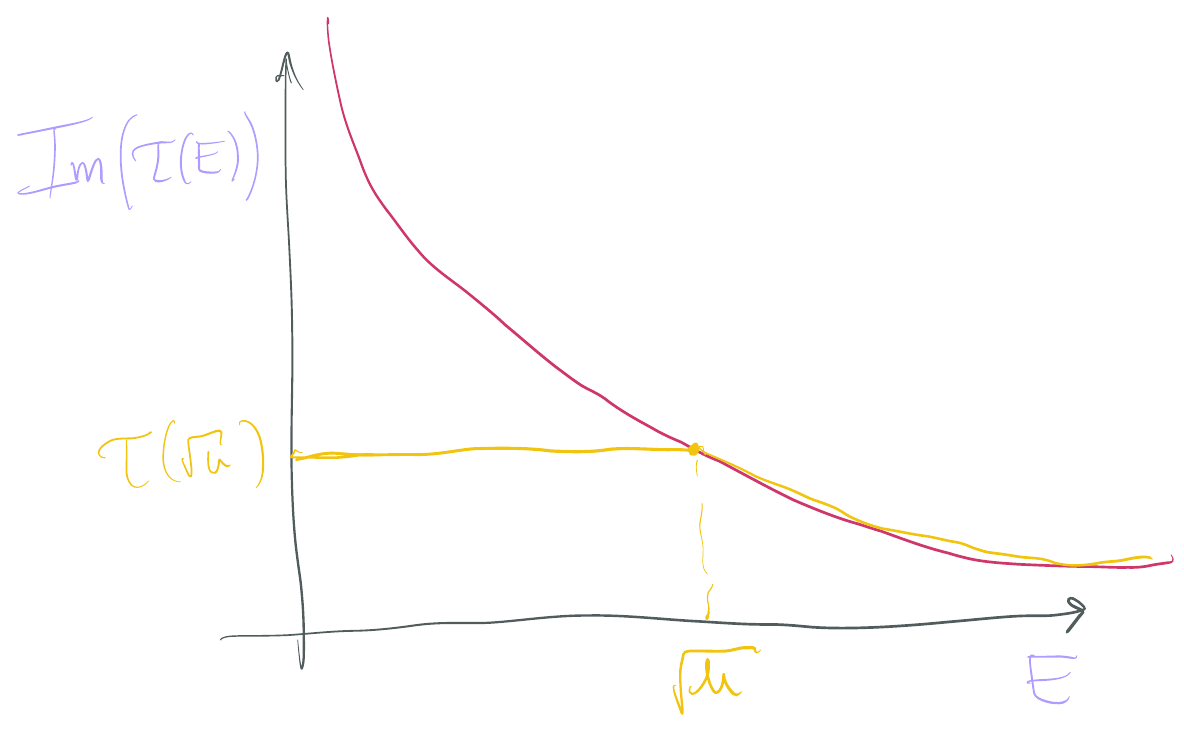}
\caption{\label{RGtau}RG-running of the holomorphic gauge coupling. $\tau(u)$ is not a constant function of the CB parameter.}
\label{MS}
\end{center}
\end{figure}

Now let us study the RG running at a generic point $u$ of $\cC$. First recall that for an $\suf(2)$ theory 
\beq\label{asu2}
a=\sqrt{u/2}
\eeq
and in particular $u$ has scaling dimension 2 \eqref{uCB}. Then notice that for energy scales $E\gg |\sqrt{u}|$, the theory is effectively an $\suf(2)$ theory and the running is governed by \eqref{oneloop}\footnote{In making this statement one has to be careful. In fact IR-free theories are not UV complete and thus are not valid at arbitrarly high energy scales. We are here implicitly assuming that at an energy $\tilde{E}<|\tilde{\L}|$ another description ``kicks in''  completing the theory in the UV. With this understanding, everything else follows.}. $|\sqrt{u}|$ is physically the energy scale at which the $\suf(2)\to U(1)$. Since the superpotential terms \eqref{SupTer} induce a mass to the hypermultiplets $m_{\rm CB}\sim |\sqrt{u}|$, for energy scales below $|\sqrt{u}|$, the theory is, as we already discussed, a pure $\cN=2$ theory of vector multiplets. In particular there are no degrees of freedom which are charged under the $U(1)$ gauge group generating non-trivial quantum corrections. At any point $u\in \cC$, the renormalization of the holomorphic coupling ``stops'' at |$\sqrt{u}$ and therefore we can obtain the perturbative contribution to the low-energy effective coupling by plugging $\mu\mapsto\sqrt{u}$ in \eqref{oneloop}:
\beq\label{tauIR}
\tau(u)=-\frac{b}{2\pi i}\log\left(\frac{\sqrt{u}}{\tilde{\Lambda}}\right)
\eeq 
This behavior is depicted in figure \ref{RGtau}. To get the full expression of the low-energy effective coupling as a function of $u$ we need to also include non-perturbative corrections. But for the moment we choose to neglect them and instead only consider \eqref{tauIR} for values of $u$ close to the origin of $\cC$, more precisely $|\sqrt{u}|\ll |\tilde{\L}|$, where non-perturbative corrections can be neglected.

From \eqref{adual} and \eqref{asu2}, we can integrate \eqref{tauIR} and obtain:
\beq\label{adloop}
a^D= -b\frac{\sqrt{u/2}}{2\pi i}\left[\log\left(\frac{\sqrt{u/2}}{\tilde\L}\right)-1\right]+{\rm non\shm pert.\ contr.}
\eeq
where $b$ is the coefficient of the gauge coupling beta function and the logarithm term corrects the classical part which is indeed linear in $a\sim\sqrt{u}$. The initial goal of the CB analysis was to exactly compute the non-perturbative corrections in \eqref{adloop} \cite{Seiberg:1994rs,Seiberg:1994aj}, we will instead neglect them by considering the theory in a regime where these corrections are un-important and focus instead in highlighting the power of the CB geometry in constraining what $\cN=2$ supersymmetric field theories are allowed.

\subsection{Multiple Lagrangian descriptions}

We have just derived the explicit expression for the special coordinates. Let's now focus on a specific example choosing an $\suf(2)$ theory with exactly five hypermultiplets in the ${\bf 2}$. For this theory $b=-1$. In section \ref{locglo} we warned the reader that the coordinate $a$ might be only locally well-defined, justifying this statement by the speculative possibility that $a$ could pick up a non-trivial transformation if dragged along a closed loop, this would lead to the surprising result that as we move around the CB, the effective low-energy description of the family of $U(1)$ theories is not unique. We will show explicitly that this is precisely what happens in examples, even in the simplest $\suf(2)$ case. The conclusion is rather bizarre and surprising; despite the extreme simplicity of the low-energy theory on $\cC$ (just a single $\cN=2$ vector multiplet!) there is no globally defined lagrangian description for it!

Let's recall again the expression for the special coordinates:
\begin{align}\label{a1}
a&=\frac12\sqrt{u/2}\\\label{a2}
a^D&=\frac{\sqrt{u/2}}{2\pi i}\left[\log\left(\frac{\sqrt{u/2}}{\tilde\L}\right)-1\right]
\end{align}
we follow \cite{Seiberg:1994aj} choosing a different normalization for \eqref{asu2}, this is appropriate for $\suf(2)$ theories with hypermultiplets in the ${\bf 2}$ to ensure that the electric charges on $\cC$ are always integral.

\begin{figure}
\begin{center}
\includegraphics[width=.4\textwidth]{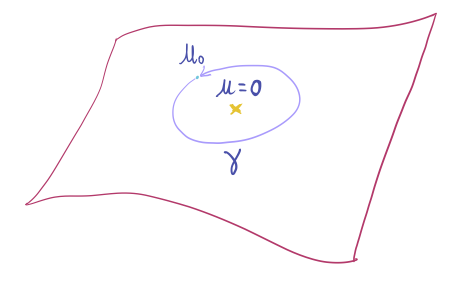}
\caption{\label{loop}Moving along a small loop around $u=0$ we return to a different description of the low-energy physics.}
\end{center}
\end{figure}

Consider now taking a loop $\g$ around the origin $\cC$ with radius $R\ll|\tilde\L|$ (so that we are in the correct regime to use \eqref{a1} and \eqref{a2}), see figure \ref{loop}. Something surprising happen to the special coordinates. As $u_0\to e^{i2\pi} u_0$, $a$ and $a^D$ don't come back to their original values but rather:
\beq\label{Ttrag}
\quad \left(\begin{array}{c}
a^D\\
a
\end{array}\right)\xrightarrow{\rm along \ \g}\left(
\begin{array}{cc}
\shm1&\shm1\\
0&\shm1
\end{array}
\right)
\left(\begin{array}{c}
a^D\\
a
\end{array}\right),\qquad M_\g:=\left(\begin{array}{cc}
\shm1&\shm1\\
0&\shm1
\end{array}
\right).
\eeq
This matrix picked up by the special coordinates is called the \emph{monodromy} along $\g$. $M_\g$ has a special from (for instance is clearly integer valued). Using \eqref{adual} we conclude that looping along $\g$ we return to a description of the theory with a different holomorphic coupling:
\beq
\t\xrightarrow{\rm along \ \g} \t+1.
\eeq
This transformation has a special name, it is called a \emph{T transformation}, or rather the negative of it, and will be discussed again in the next lecture. Is this behavior consistent with what we know about $\cN=2$ supersymmetric gauge theories? You are probably intuitive enough to guess that this is a rhetorical question and the answer if of course yes.  The resolution of this puzzle will lead us to realize something possibly even stranger; there is no global lagrangian description of the low-energy effective theory of a $\cN=2$ supersymmetric theory on its CB and many different, yet equivalent, descriptions of the theory of $\cN=2$ $U(1)$ gauge theories are needed. And looping around closed loops we come back to a description that is different but it is equivalent to the initial one. This set of equivalent descriptions is called \emph{electromagnetic duality} and it will describe in detail in the next section. 


Another important observation on how multiple lagrangian descriptions of the low-energy theory at a point $u_0$ of $\cC$ could arise is to notice that, as we showed in exercise \ref{KaMetU1}, at the origin of $\cC$ the K\"ahler metric diverges. Thus as a metric space, the regular points $\cC_{\rm reg}=\C/\{0\}$ and the loop $\g$ which we considered encircles the singular point. The non-triviality of $M_\g$ arises precisely because $\g$ encircles the origin. The existence of singularities on the CB of $\cN=2$ theories is a very important concept to absorb and deserves a bit more discussion.

\subsection{Singularities on the CB}

The K\"ahler metric on $\cC$ \eqref{metric} is determined by Im$\tau(u)$. As we discussed, we generically expect that if the beta function does not vanish, $\t$ runs and therefore picks a non-trivial dependence on $u$. In addition both its physical interpretation as the holomorphic gauge coupling of the $U(1)$ theory and the positivity of the metric, bound Im$\tau>0$. But if Im$\tau$ (as well as Re$\tau$) is a bounded harmonic function on $\R^2$ (real and imaginary part of $u$), it must be a constant function over the entire CB. How is this compatible with the RG-running of $\t$ and its non trivial dependence on $u$?

This contradiction was noticed in \cite{Seiberg:1994rs} starting in the case of $\cN=2$ $\suf(2)$ Yang-Mills theory;\footnote{This theory is asymptotically free. In the description of \cite{Seiberg:1994rs}, the perturbative calculation of the special coordinates could only be trusted at very high energy. Thus the analysis in some ways opposite to ours as the physics which is hard to calculate happens near the origin and their reasoning involves taking a loop $\g$ with a very large radius.} and was resolved by realizing that the CB is not $\C$ but rather $\cC=\C\setminus\cV$, with $\cV$ the set of singularities. In fact the existence of singularities is not just an accident  but it has  deep implications on the low-energy phsyics on $\cC$. But what is the physical interpretation of such singularities?

\begin{tcolorbox}
\begin{center}
\textbf{Coulomb branch singularities}
\end{center}\vspace{-.5em}
The relations \eqref{metric} show that the metric structure on $\cC$ is equivalent, in the special coordinates basis, to the holomorphic gauge couplings of the low-energy effective description of our theory. The existence of a singularity in the metric, signals that there something has gone wrong with our effective-action description. What happens if there are values of $u$ for which the effective $U(1)$ low-energy theory has extra $U(1)$ charged degrees of freedom which become massless? The fact that those degrees of freedom are charged under the low-energy $U(1)$, implies that integrating them out affect the behavior of the theory in the IR. The fact that they are massless, tells us that we are no-longer justified to integrate them out.
\end{tcolorbox}\vspace{0.5em}

\begin{dig}[]{}
Since $\cC$ is both a complex and metric space, singularities can and do arise in both structures. In this series of lectures we will strictly \emph{metric singularities}. As a complex manifold, the CB of the $\suf(2)$ theories we discussed has no singular behavior at the origin, $\cC\cong \C$. This has important physical consequences as complex singularities on the moduli space have a very different physical interpretation \cite{Argyres:2017tmj,Argyres:2018wxu}.
\end{dig}

\begin{ex}[]{}
\begin{exe}\label{Renorm}
Using the standard result of the one-loop beta function 
\beq\label{Oneloop}
b=\frac{11}{6}T({\rm adj})-\frac13\sum_fT({\bf R}_f)-\frac16\sum_bT({\bf R}_b)
\eeq
show that for $\cN=1$ susy gauge theory $b$ reduces to
\beq
b=\frac{3}{2}T({\rm adj})-\frac{1}{2}\sum_cT({\bf R}_c)
\eeq
where the sum runs over the number of chiral multiplets. Using $T({\bf R})=T(\bar{{\bf R}})$, show that for $\cN=2$ susy gauge theories
\beq\label{N2bfun}
b=T({\rm adj})-\sum_hT({\bf R}_h)
\eeq
where the sum runs over the representation of the hypermultiplets.
\end{exe}

\begin{center}
\rule[1mm]{2cm}{.4pt}\hspace{1cm}$\circ$\hspace{1cm} \rule[1mm]{2cm}{.4pt}
\end{center}

\begin{exe}\label{KaMetU1}
Use the quantum corrected expression for $a^D$ \eqref{adloop} and compute the K\"ahler metric on $\cC$. Show that the metric induced on the $\cC$ is classically constant but it is quantum mechanically singular as $u\to 0$.
\end{exe}

\begin{center}
\rule[1mm]{2cm}{.4pt}\hspace{1cm}$\circ$\hspace{1cm} \rule[1mm]{2cm}{.4pt}
\end{center}

\begin{exe}\label{MonodromySU2}
Show that for an (IR-free) $\suf(2)$ theory with $n+4$ hypermultiplets in the ${\bf 2}$, the monodromy around $u=0$ is:
\beq\label{MonSu2}
M_{\gamma}=-T^n
\eeq
where $n>0$ is the negative of the coefficient of the beta function.
\end{exe}

\begin{center}
\rule[1mm]{2cm}{.4pt}\hspace{1cm}$\circ$\hspace{1cm} \rule[1mm]{2cm}{.4pt}
\end{center}

\begin{exe}\label{MonodromyU1}
Using the fact that for a $U(1)$ $\cN=2$ theory the coefficient of the beta function is:
\begin{align}\label{U1bfun}
U(1)&:\qquad b=\sum_i^n \cQ^2_i
\end{align}
show that for a $U(1)$ theory the monodromy around the $u=0$ is instead:
\beq\label{MonSu2}
M_{\gamma}=T^n
\eeq
(Hint: what's the relation between $u$ and $a$ for $U(1)$ theories?)
\end{exe}

\begin{center}
\rule[1mm]{2cm}{.4pt}\hspace{1cm}$\circ$\hspace{1cm} \rule[1mm]{2cm}{.4pt}
\end{center}

\begin{exe}\label{Mass}
Consider the $\suf(2)$ theory with 6 hypermultiplet in the ${\bf 2}$ that we have analyzed in the previous section. We showed that it has one singularity at the origin of $\cC$. Identify which charged states become massless there. 
\end{exe}
\end{ex}


\section{Electro-Magnetic Duality}\label{EMduality}
\setcounter{exe}{0}

In order to make sense of the seemingly inconsistent story which we have just derived, we need to discuss yet another property of a low-energy $\cN=2$ $U(1)$ theory. In particular we will now show that there are many different, ``dual'', descriptions of the same theory. Thus is physically consistent to have a non-globally defined description so long as it involves dual ones. The example we studied is precisely of this kind. In order to understand how this works we need to also include magnetic monopoles.

Consider first a non-supersymemtric $U(1)$ gauge theory (or else restrict to the the gauge field terms in
the bosonic part of the action). Working in the Minkowski space with
conventions $(F_{\mu\nu})^2=-(\widetilde F_{\mu \nu})^2$ and
$\widetilde{\widetilde F}=-F$, these
terms can be written as 
\beq\label{Lagin}
\frac{1}{32 \pi} \, {\rm Im}\, \int \, \tau (a) (F+i\widetilde F)^2 =
\frac{1}{16 \pi} \, {\rm Im} \, \int \, \tau (a) (F^{2}+i\widetilde
FF) \,.
\eeq
Now we regard $F$ as an independent field and implement the Bianchi 
identity $dF = 0$ by introducing a Lagrange multiplier vector field
$A^D_\mu$. To fix the Lagrange multiplier term, $U(1)\subset SU(2)$ is
normalized such that all $SO(3)$ fields have integer charges. Then all matter fields in the fundamental
representation of $SU(2)$ will have half-integer charges. With this 
convention, a magnetic monopole satisfies $\epsilon^{0\mu\nu\rho}
\partial_{\mu}F_{\nu\rho}=8\pi\delta^{(3)}(x)$. The Lagrange
multiplier term can now be constructed by coupling $A^D_\mu$ to a
monopole:
\beq\label{LagMult}
\frac{1}{8\pi}\,\int\,A^D_\mu\epsilon^{\mu\nu\rho\sigma}
\partial_{\nu}F_{\rho\sigma}=\frac{1}{8\pi}\,\int\,\widetilde F_{D}F = 
\frac{1}{16\pi}\,{\rm Re}\,\int\,(\widetilde
F_{D}-iF_{D})(F+i\widetilde F)\,,
\eeq
where, $F^D_{\mu\nu}=\partial_\mu A^D_\nu-\partial_\nu A^D_\mu$.

Consider then the following lagrangian
\beq\label{Lag1}
\cL_{\rm EM}\sim{\rm Im}\int \frac{\tau}{2}\left(F+i\tilde{F}\right)^2+2\int \tilde F^D F.
\eeq
Integrating out $F^D$ \eqref{Lag1}, by constructions, correctly reproduces Maxwell's equations in the absence of charges. We can now treat $F^D$ as a dynamical field and instead integrate out $F$. Performing this calculation we obtain
\beq\label{LagDua}
\cL^D_{EM}\sim \frac{1}{2}{\rm Im}\int -\frac{1}{\tau}\left(F^D+i\tilde{F}^D\right)^2
\eeq
which is a description of the same theory but instead in terms of the dual variable which, by construction, couples to monopoles rather than electrically charged particles. \eqref{LagDua} describes a famous duality in electromagnetism called \emph{S-duality} \cite{Montonen:1977sn,Cardy:1981qy,Cardy:1981fd,Shapere:1988zv} and shows that the physics of electromagnetism is invariant under the following transformation:
\beq\label{Sdual1}
A_\mu\to A_\mu^D\quad{\rm as\ well\ as} \quad \tau\to-\frac{1}{\tau}
\eeq

All the steps which lead to \eqref{Sdual1} can also be performed in an $\cN=1$ supersymmetric
language finding an analogous result. In the case of the $\cN=2$ supersymmetric lagrangian, the dual description which we just found relate the scalars $\sCB$ and $\sCB^D$ and their vevs. Recall that on a generic point of the Coulomb branch
\beq
\tau(a)=\frac{\partial a_D}{\partial a}
\eeq
from which we can write the S-duality transformation in $\cN=2$ language:
\beq\label{Sdual}
\left(\begin{array}{c}
a^D\\
a
\end{array}\right)\to\left(
\begin{array}{cc}
0&1\\
\shm1&0
\end{array}
\right)
\left(\begin{array}{c}
a^D\\
a
\end{array}\right)\qquad
\&\qquad
\tau\to-\frac{1}{\tau}
\eeq
this transformation is also called \emph{S tranformation}.

But \eqref{Sdual} does not exhaust all the transformations which leave the physics of a $\U(1)$ $\cN=2$ theory invariant. In fact from the definition of the holomorphic gauge coupling, shifting $\th\to\th+2\pi$ provides another transformation which gives an equivalent description of the same physical system:
\beq\label{Ttra}
\left(\begin{array}{c}
a^D\\
a
\end{array}\right)\to\left(
\begin{array}{cc}
1&1\\
0&1
\end{array}
\right)
\left(\begin{array}{c}
a^D\\
a
\end{array}\right)\qquad
\&\qquad
\tau\to\tau+1
\eeq
this transformation, which the reader will recognize to be related to $M_\g$ in \eqref{Ttrag}, is instead called a \emph{T transformation}. Combining \eqref{Ttra} and \eqref{Sdual} we can generate an infinite, albeit discrete, set of transformations which are parametrized by $SL(2,\Z)$. The latter is therefore identified
with the full duality group of our theory. This group acts on $\tau$ and the special coordinates as follows:
\beq
\left(\begin{array}{c}
a^D\\
a
\end{array}\right)\to\left(
\begin{array}{cc}
a&b\\
c&d
\end{array}
\right)
\left(\begin{array}{c}
a^D\\
a
\end{array}\right),\qquad
\&\qquad
\tau\rightarrow{\frac{a\tau + b}{c\tau + d}}\,,
\label{tauSLtwoZ}
\eeq
where, $ad-bc=1$ and $a,b,c,d \in \Z$. It is important to clarify that $\SL(2,\Z)$ it is \textbf{\emph{not}} a symmetry of the theory. It in fact acts on the coupling constant and that is why we refer to it as a duality group. It maps one description of the theory into a different but physically equivalent one.\footnote{There exists special values of the holomorphic gauge couplings which are fixed by finite subgroups of $\SL(2,\Z)$. For those values of $\t$ and these finite subgroups do act as honest symmetry of the theory.}

For an $\cN=2$ theory of rank $r$, the EM duality group is instead $\Sp(2r,\Z)$. A convenient way of parametrize this group is:
\beq\label{Sp2rZ1}
      \Sp(2r,\Z)\ni M=\left(
       \begin{array}{c;{2pt/2pt}c}
        A & B \\ \hdashline[2pt/2pt]
        C & D 
    \end{array}\right)\qquad M^T J M=J
    \eeq

where $A$, $B$, $C$ and $D$ are $n$ by $n$ matrices and $J$ is a $2n$ by $2n$ non-degenerate skew-symmetric matrix. If we make the following choice:
\beq\label{Sp2rZ2}
J=\left(
 \begin{array}{c;{2pt/2pt}c}
        0 & \I_n \\ \hdashline[2pt/2pt]
        -\I_n & 0 
    \end{array}
\right)
\eeq
The group action on the special coordinates and the holomorphic gauge coupling straightforwardly generalizes \eqref{tauSLtwoZ}:
\beq
\quad A^TD-C^TB=\I_n,\quad A^T C\quad{\rm and}\quad B^TD\quad{\rm symmetric}\quad
\eeq
then \eqref{tauSLtwoZ} generalizes straightforwardly:
\beq
\left(\begin{array}{c}
{\boldsymbol a}^D\\
{\boldsymbol a}
\end{array}\right)\to\left(
\begin{array}{cc}
A&B\\
C&D
\end{array}
\right)
\left(\begin{array}{c}
{\boldsymbol a}^D\\
{\boldsymbol a}
\end{array}\right),\qquad
\&\qquad
{\boldsymbol\tau}\rightarrow{\frac{A{\boldsymbol \tau} + B}{C{\boldsymbol \tau} + D}}\,,
\label{tauSp2rZ}
\eeq
here ${\boldsymbol a}^D$ and ${\boldsymbol a}$ are both $r$-component vectors and ${\boldsymbol \tau}=\partial {\boldsymbol a}^D/\partial {\boldsymbol a}$.

In light of what we learned about electro-magnetic duality, it is physically consistent if the special coordinates are not single valued on $\cC$, so long as the different value are related by an element of the low-energy EM duality group. This is precisely what happens in \eqref{Ttrag} and what then gives a consistent description of the low-energy physics of theory we analyzed in the previous section. It also shows another important fact, the special coordinates are not holomorphic \emph{functions} on $\cC$ but rather a holomorphic 
\emph{section} of an $\SL(2,\Z)$ bundle. The structure group of the bundle over $\cC$ in the rank-$r$ case is $\Sp(2r,\Z)$ instead.

\begin{ex}[]{}
\begin{exe}\label{Monopole}
Consider the following which is obtained by adding \eqref{LagMult} to \eqref{Lagin} and a coupling term $k^\mu A^D_\mu$:
\beq\label{LagTot}
\frac{1}{32 \pi} \, {\rm Im}\, \int \, \tau (a) (F+i\widetilde F)^2 +\frac{1}{16\pi}\,{\rm Re}\,\int\,(\widetilde
F_{D}-iF_{D})(F+i\widetilde F)+ \int k^\mu A^D_\mu
\eeq
Check that integrating out the dual $A^D_\mu$ from \eqref{LagTot} reproduces the correct Maxwell's equations in the presence of a magnetic current\label{VI}:
\beq
\epsilon^{\mu\nu\rho\s}\partial_\nu F_{\rho\s}=k^\mu
\eeq
\end{exe}

\begin{center}
\rule[1mm]{2cm}{.4pt}\hspace{1cm}$\circ$\hspace{1cm} \rule[1mm]{2cm}{.4pt}
\end{center}

\begin{exe}\label{dual}
Integrate out $F$ from \eqref{Lag1} and check \eqref{LagDua}.
\end{exe}

\begin{center}
\rule[1mm]{2cm}{.4pt}\hspace{1cm}$\circ$\hspace{1cm} \rule[1mm]{2cm}{.4pt}
\end{center}

\begin{exe}\label{ShriLoop}
Now that we have shown that physical consistency implies that the monodromies $M_\g$ have to be valued in the discrete group $\SL(2,\Z)$, think of a simple topological argument to show that a monodromy along a loop $\g$ can only be non-trivial if the loop encircles at least one singularity.
\end{exe}
\end{ex}


\section{Rank-1 scale invariant case}\label{sec5}
\setcounter{exe}{0}

It is useful to summarize the properties that we have thus far ``discovered'' and which characterize the CB of a rank-1 $\cN=2$ gauge theory:

\begin{tcolorbox}
\begin{center}
\textbf{Summary of Coulomb branch geometry}
\end{center}\vspace{-.5em}
\begin{itemize}[leftmargin=*]
\item[\textbf{\dred{1.}}] \textbf{The existence of a one complex dimensional space $\cC$, the actual \emph{Coulomb branch}}. This space parametrizes the allowed gauge inequivalent vacuum configurations of the scalar component of the $\cN=2$ vector multiplet. In each vacuum the theory in the deep IR is effectively a $U(1)$ theory with no charged degrees of freedom.

\item[\textbf{\dred{2.}}] \textbf{The existence of a holomorphic section of a two dimensional (flat) $SL(2,\mathbb{Z})$ vector bundle $\left(\begin{array}{c}a^D(u)\\a(u)\end{array}\right)$ from which we can compute the holomorphic gauge coupling of the low-energy $U(1)$ theory: $\tau(u)=\partial a^D(u)/a(u)$}. The $a^D$ and $a$ are called the \emph{special coordinates} on $\cC$ and they allow to extract important physical information from the geometric data on $\cC$. $a^D$ is also interpreted as the vacuum expectation value of the $\cN=2$ scalar partner of the dual photon $A^D_\mu$.

\item[\textbf{\dred{3.}}] \textbf{The existence of a \emph{K\"ahler} metric on $\cC$}. The metric, when expressed in terms of the \emph{special coordinates}, is just given by $ds^2={\rm Im} \tau d \bar{a}\otimes da$, where $\tau$ is holomorphic gauge coupling of the the low-energy $U(1)$ theory and its imaginary part is therefore positive definite.

\item[\textbf{\dred{4.}}] \textbf{The existence of metric singularities characterized by non trivial $SL(2,\Z)$ monodromies}. Because of these metric singularities, $\cC$ is not a K\"ahler manifold, $a^D$ and $a$ pick up a non-trivial monodromy when dragged along a closed loop encircling the singularity.
\end{itemize}
\end{tcolorbox}\vspace{0.5em}

All combined these properties define what we will call \emph{Coulomb branch geometry} or \emph{Seiberg-Witten geometry}. This geometric structure is also known in the mathematics literature as Special K\"ahler geometry, though often in the math literature the electromagnetic duality group is defined over the reals rather than the integers \cite{Freed:1997dp}.\footnote{The fact that the electromagnetic duality group is constrained to be over the integers is a result of the fact that $\cN=2$ theories have an extra structure that we have not discussed. This is the lattice of charges $\L$ which captures the charges of the BPS states of the physical theory. We will not discuss this point any further here.}

\begin{definition}[Singular (rank-1) Special K\"ahler space]{SpecialKa}
We define a Singular Special K\"ahler space as a triple $(\cC,\sigma,\tau)$ where:\
\begin{itemize}
\item $\cC$ is a K\"ahler space of complex dimension 1.

\item $\sigma(u):=\left(\begin{array}{c}a^D(u)\\a(u)\end{array}\right)$ is a two dimensional section of an $SL(2,\Z)$ bundle over $\cC$.

\item $\tau:=\frac{\partial a^D}{\partial a}$ is such that Im$(\tau)>0$ and $ds^2=({\rm Im}\tau) d\bar{a}\otimes da$.
\end{itemize}
\end{definition}

\subsection{Possible rank-1 $\cN=2$ SCFTs}

To show the power of this geometric structure we can try to ask an ambitious question: \emph{can we use our machinery to list all rank-1 CB geometries which are compatible with $\cN=2$ superconformal invariance?} In carrying out this task, we will philosophically take the geometry of $\cC$ as more fundamental than the $\cN=2$ theory which has $\cC$ as its CB and let the former constrain the latter. In this way the construction of CB geometries which are compatible with the Special K\"ahler requirements can teach us about the existence of the theories which we otherwise would not know they existed. In the following we will use the letter $\cT$ to indicate a generic four dimensional $\cN=2$ superconformal field theory.

\begin{figure}
\begin{center}
\includegraphics[width=.8\textwidth]{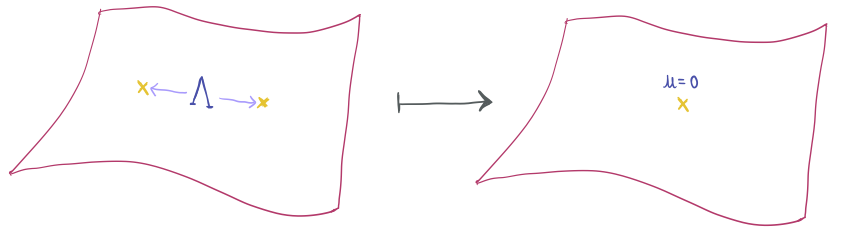}
\caption{\label{scaInv}The existance of more than one singularities at finite metric distance breaks scale invariance (left). Therefore the singularity structure of rank-1 CBs is tightly constrained (right).}
\label{SInv}
\end{center}
\end{figure}

The way we will encode conformal invariance in our geometrical data, is to require that our CB geometry is scale invariant. Immediately this requirement constraints dramatically the number of singularities which can appear. In fact, as depicted in figure \ref{scaInv}, scale invariance readily implies that we can only have one singularity at finite distance located at the origin of $\cC$. Scale invariance does not forbid a singularity at infinity and in fact we will shortly see that the geometries we construct also have a singularity at infinity.

Let's be more specific in defining this scaling action and its relation with superconformal invariance. Operators in a quantum theory are organized in representations of the symmetry algebra which can be effectively label by the eigenvalues of maximally commuting set of its generators. For a superconformal theory $\cT$ the eigenvalues corresponding of this maximal commuting set of operators are:\footnote{For a more systematic discussion of superconformal invariance and the constraints on the operator algebra of a superconformal field theory see Lorenz and Madalena's notes.}
\begin{align}
R\ &:\ \suf(2)_R\\
r\ &:\ \U(1)_r\\
[j_1;j_2]\ &:\ \sof(3,1)_{\rm Lorentz}\cong \suf(2)_L\times \suf(2)_R\\
\D\ &:\ \R^+\text{corresponding to scaling}
\end{align}
and we will use the following notation to label quantum numbers of the operators \cite{Cordova:2016emh}:
\beq
\cO[j_1;j_2]^{(R_\cO;r_\cO)}_{\D_\cO}
\eeq
We will call $\D_\cO$ the \emph{scaling dimension} of the operator $\cO$.


Superconformal invariance of $\cT$, plus consistency conditions of OPE coefficients with vacuum expectation value \cite{Argyres:2019yyb}, imply that the scalars which acquire a vev, and which parametrize the different branches of the moduli space of vacua, have to be superconformal primaries of specific (short) multiplets in $\cT$\footnote{For a proper definition of the notion of a conformal and superconformal descendent, as well as a classification of the $\cN=2$ superconformal multiplets in four dimension see Lorenz's notes.}. For example some of the multiplets that we have already encountered can written as:
\beq\label{N2multiplets}
\def\arraystretch{1.2}
\begin{array}{c|c}
\textsc{Field}&\quad \cN=2\ \textsc{Multiplet \cite{Cordova:2016emh}}\\
\hline
\hline
\quad\cN=2 \ {\rm vector\ multiplet}\quad{}& A_2\bar{B}_1[0;0]^{(0;2)}_1 \\
\quad\cN=2\ {\rm Coulomb\ branch\ operator}\ (\text{e.g. Tr}\big[\Phi^n\big])\quad{} &L\bar{B}_1[0;0]^{(0;2r)}_r \\
\end{array}
\eeq
Therefore we can identify the CB coordinate $u$ as the vev of \eqref{N2multiplets}:
\beq\label{Cbu}
u:=\langle L\bar{B}_1[0;0]^{(0;2\D_u)}_{\D_u}\big|_{\rm sp}\rangle
\eeq
where the subscript $|_{\rm sp}$ simply specifies that it is only the superconformal primary of the multiplet which acquires a vev.  In the following we will often leave this implicit.  We then notice that the operators which acquire a vev on the CB have the perculiar property that their $\U(1)_r$ charge is proportional to their scaling dimension $\D$, since $\D_\cO\geq 1$ from unitarity constraints, they are both also non zero. On a generic point of $\cC$, the $\U(1)_r\times \R^+$ is spontaneously broken and their combined action gives rise to a $\C^*$ action on $\cC$ where the weight of this action is proportional to the operators' scaling dimension:\footnote{A generalization of this $C^*$ action, which also include the action of the Cartan of the $\suf(2)_R$ symmetry, can also be defined on the entire moduli space $\cM$}
\beq\label{ScaAct}
\C^*\ni \lambda \circ \langle \cO\rangle:=\l^{\D_{\cO}}\langle \cO\rangle
\eeq

The special coordinates are also identified with expectation values of the scalar component of the (gauge invariant) $\cN=2$ vector multiplet associated to the free photon and dual photon on a generic point of the CB. For the following we are only interested in the quantum numbers of the corresponding operators, which in the case of $a^D$ and $a$ are the same. We will thus not make a distinction between the operators corresponding to the special coordinates though the reader should be clear that $a^D$ and $a$ should in fact be identified as vevs of different operators in the IR. From \eqref{N2multiplets}:
\beq\label{SpecOpe}
a^D \& \ a:=\langle A_2\bar{B}_1[0;0]^{(0;2)}_1\big|_{\rm sp}\rangle
\eeq

The scaling action \eqref{ScaAct} applied to the triple $(u,a^D,a)$ will then give:
\begin{align}\label{Csrank1}
\l\,\circ:\left\{
\begin{array}{l}
 u\\
 a^D\\
 a
\end{array}
\right.
\quad\mapsto\quad 
\begin{array}{l}
\lambda^{\D_u}\ u\\
\lambda\ a^D\\
\lambda\  a
\end{array} 
\quad \l \in \C^*,
\end{align}
here $\D_u$ is the scaling dimension of $u$ which, by unitarity bounds for 4d CFTs satisfies $\D_u \geq 1$\footnote{Here we are throwing some subtleties under the rag. Strictly speaking this implication is only valid if the coordinate ring of the CB is freely generated as we are implicitly assuming throughout these lectures. If not, apparent violation of the unitarity bound might arise \cite{Argyres:2017tmj}. These are consequences of non-trivial relations among CB operators.}. Recall that the $\C^*$ action is not a general property of CB geometries and it only arises in the superconformal. In which case all the properties of the CB must respect this action and we will say that the CB is \emph{scale invariant} and its corresponding geometry a \emph{scale invariant special Kahler geometry}. If $\cC$ is scale-invariant, there is a single value of the CB coordinate $u$ which is left invariant under the scaling action \eqref{Csrank1} and it is the origin $u=0$ of $\cC$. As we mentioned in passing above, we expect this point to give rise to a metric singularity and we will often call it the \emph{superconformal vacuum} since it is where our superconformal field theory $\cT$ ``lives''.

\begin{figure}
\begin{center}
\includegraphics[width=.5\textwidth]{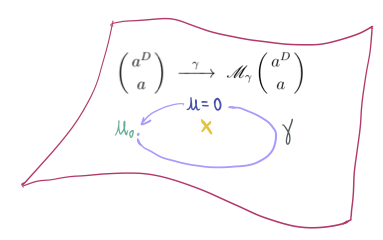}
\caption{\label{Loop} By moving around the singularity $u=0$, where the $\cN=2$ SCFT $\cT$ sits, the special coordinate pick-up a monodromy transformation $\mathscr{M}_\g$.}
\label{loop}
\end{center}
\end{figure}

To show the power of scale invariant special Kahler geometry, we want to now impose the conditions which we summarized at the beginning of this section and ask systematically how many of these spaces can we construct. We will find, perhaps to the surprise of the reader, that special K\"ahler constraints are tight enough to allow only a finite set of possibilities. It is important to remark once more that in carrying out this calculation we are abstracting from the comforting lagrangian framework. This means that we are lifting any pre-conceded relation between $a$ and $u$ and in particular we are not assuming that the complex coordinate $u$ on $\cC$ is identified with $a^2$. The scaling dimension $\D_u$ will be therefore an outcome of this discussion and so the functional dependence of $a(u)$. 

The main object which we will leverage to complete our analysis will be the $\SL(2,\Z)$ monodromy. This matrix is associated to those closed loops in $\cCrg$ which cannot be shrunk to a point. This follows from the fact that the monodromy takes values over the integers and admits no non-trivial continuous dependence on a parameter is possible. So what determines the particular $\SL(2,\Z)$ value of the matrix is the homology class of a loop $[\g]$ rather than the loop itself.    

We have already argued that scale invariant CBs have a single singularity at the origin; $\cCrg \simeq \C\backslash \{0\}$, therefore their fundamental group is $\pi_1(\cCrg)\cong \Z$ is generated by a path $\g$ that circles once around $u=0$ counterclockwise. We call $\mathscr{M}_\gamma \in \SL(2,\Z)$ the monodromy associated to this loop which will act on the special coordinates as in \eqref{Csrank1} once we drag them along $\g$, see figure \ref{Loop}. To constrain the allowed values for $\Msc_\g$ we can use a trick by noticing this $\g$ can be see as an orbit of the $\C^*$ action \eqref{Csrank1} which naturally acts on $\cC$ and the $(a,a^D)$. Specifically:
\beq\label{loop}
\g:=\{\lambda(t)\circ u_0\ |\ \lambda(t):=e^{i2\pi t/\D_u},\ t\in[0,1]\}
\eeq 

From which we can derive the following constrain on the way $(a,a^D)$ transform after looping around $\g$:
\beq
\left(
\begin{array}{c}
a^D\\
a
\end{array}
\right)\ \xrightarrow{\ \ \g\ \ }\ e^{i2\pi/\D_u}\left(
\begin{array}{c}
a^D\\
a
\end{array}
\right)\equiv\ \mathscr{M}_\g
\left(
\begin{array}{c}
a^D\\
a
\end{array}
\right)
\eeq
Two important fact immediately follow:\\

\begin{tcolorbox}
\begin{center}
\textbf{Conditions for rank-1 monodromies}
\end{center}\vspace{-.8em}
\begin{itemize}[leftmargin=*]
\item[$i$)] The special coordinates must be proportional to an eigenvector of $\mathscr{M}_\g$.
\item[$ii$)] $e^{i2\pi/\D_u}$ has to be an allowed eigenvalue for an $\SL(2,\Z)$ matrix.   Therefore, $\mathscr{M}_\g$ must have an eigenvalue, $\m=\exp(2\pi i/\D_u)$, with $|\m| = 1$. 
\end{itemize}
\end{tcolorbox}\vspace{0.5em}

From exercise \ref{SL(2,Z)} we find that if $\mathscr{M}_\g\in\SL(2,\Z)$ there are only a finite set of eigenvalues satisfying these conditions:  
\beq\label{aleig}
\mu=\pm1,\ e^{\pm i \pi/3},\ e^{\pm i \pi/2},\ e^{\pm i 2\pi/3}.
\eeq
and since $\D_u\ge1$ we can immediately read off the list of allowed values of $\D_u$ and $\mathscr{M}_\g$ shown in Table \ref{table:eps}.

\begin{table}
\centering
\def\arraystretch{1.2}
$\begin{array}{|c|c|c|c|c|}
\hline
\ \D_u\ \,& \mathscr{M}_\g               & \t         &a^D(u) &a(u)          \\
\hline
6           & ST              & e^{i\pi/3} &e^{i\pi/3}u^{\frac16} & u^{\frac16}     \\
4           & S                & i              &i\, u^{\frac14} & u^{\frac14}       \\
3           & (-ST)^{-1} & e^{2i\pi/3} &e^{i2\pi/3}u^{\frac13} & u^{\frac13}    \\
2           & -\I               & \ \text{any}\ \t\ \,&\t \sqrt{u} & \sqrt{u}  \\
3/2        & -ST            & e^{2i\pi/3}   &e^{i2\pi/3}u^{\frac23} & u^{\frac23}   \\
4/3        & S^{-1}      & i                  &i\, u^{\frac34} & u^{\frac34}    \\
6/5        & (ST)^{-1}  & e^{i\pi/3}   &e^{i\pi/3}u^{\frac56} & u^{\frac56}    \\
1           & \I                & \text{any}\ \t &\t\,u & u \\
\hline
\end{array}$
\caption{Possible values of $\D_u$, $M$, $\t$, $a^D$ and $a$ for rank-1 CB singularities.\label{table:eps} Below the dashed line are the IR-free geometries which are discussed below.}
\end{table}

We still haven't used the full constraining power of this analysis. The eigenvector associated to the various monodromy entries in Table \ref{table:eps} are of the form:
\beq
\mathscr{M}_\g\left(
\begin{array}{c}
\xi\\
1
\end{array}
\right)=\mu\left(
\begin{array}{c}
\xi\\
1
\end{array}
\right)\qquad \xi\in\C\quad\&\quad {\rm Im}\xi>0
\eeq
From condition $i$) it then follows that the special coordinates have to have the form:
\beq
\left(
\begin{array}{c}
a^D(u)\\
a(u)
\end{array}
\right)=f(u)\left(
\begin{array}{c}
\xi\\
1
\end{array}
\right)
\eeq
Which allows us to immediately derive a general result about scale invariant rank-1 theories:

\begin{tcolorbox}
\beq
\tau(u)=\frac{\del a^D}{\del a}=\xi\quad\text{is constant}
\eeq
\end{tcolorbox}\vspace{0.5em}
To then completely solve for the special Kahler geometry and fix the $u$ dependence of the special coordinates we notice that $f(u)$ cannot be single valued around the $u=0$ and indeed:
\beq
f(e^{i2\pi}u)=\mu f(u)
\eeq
Solving this condition for all $\mu$ in \eqref{aleig} reproduces the result quoted in Table \ref{table:eps}.

Let's take a second to reflect on what we have done. First, our discussion in this section has been completely bottom-up. Systematically imposing conditions 1-4 above, we were able to, relatively effortlessly, to construct the full special K\"ahler geometry of the allowed rank-1 CBs and the full result is reported in table \ref{table:eps}. What have learned from this analysis? A first question that we might to try to answer is how many of the geometry that we found can be interpreted as CBs of superconformal gauge theories? This answer is quickly addressed by looking at the scaling dimension of $u$. There are only two gauge algebras of rank-1, $\U(1)$ and $\suf(2)$. We have discussed in previous section that $\D_u=1$ for $\U(1)$) and $\D_u=2$ for $\suf(2)$ hitting only the fourth and the bottom entry in table \ref{table:eps}. A second question that we might try to answer is what the hell do the other entries represent? And here is where the beautiful world of non-lagrangian field theories starts. What we know thus far is that if a theory with any of the entries in table \ref{table:eps} exists, it will have $\cN=2$ supersymmetry and superconformal invariance by construction.

Using a variety of methods, all the CBs that we have constructed have been realized as moduli space of vacua of $\cN=2$ superconformal field theories and some of these theories have manifested exotic properties. See for instance \cite{Argyres:1995jj,Minahan:1996fg,Minahan:1996cj}. The entries where $u$ has fractional scaling dimension correspond to the so-called rank-1 Argyres-Douglas theories \cite{Argyres:1995jj,Xie:2012hs}. Fractional CB scaling dimensions is what defines an Argyres-Douglas theory (these arise in class-S in the presence of \emph{irregular punctures} \cite{Xie:2012hs}).

\subsection{IR-free theories}

For completion, it is useful to explain how the example that we had previously studied, that is an $\cN=2$ supersymmetric $\suf(2)$ gauge theory with 5 hypermultiplets in the ${\bf 2}$, fits in this description. For that we need to understand how to accordingly modify the argument above to account for IR-free theories which are only locally (that is close enough to the origin) scale invariant.

In order to account for IR-free theories, where $\t = i\infty$, we need to also consider the $\SL(2,\Z)$ elements which are conjugate to $T$ transformations (these are the parabolic elements which we have discussed in exercise \ref{SL(2,Z)}) and have the property of only having a single eigenvalue $\pm1$ which reflects the fact that they cannot be diagonalized but only brought to a Jordan form. In this case no scale-invariant solution for the special coordinates, which we will collectively label as $\s$, exist. We should then look for solutions by including the leading corrections to scaling, e.g., expand
$\s(u) = u + \s_0 u\left(u/\L\right)^{\b_0} + \s_1 u \ln^{\b_1} \left(u/\L\right)$, where the $\b_j$ are 2-component vectors of exponents correlated with the entries of the $\s_j\in\C^2$, and $\L$ is an arbitrary mass scale.  If $\D_u =1$ (which corresponds to $\cN=2$ $U(1)$ theories)  we look for a solution to $\s(e^{2\pi i}u) = T^n \s(u)$ (see exercise \ref{MonodromyU1}).  We find $\s_0 =(0\ 0)$, $\b_1 = (1\ 0)$ and $\s_1 = ( \frac{n}{2\pi i }\ 0)$.  Thus for the $T^n$ monodromies we find
\begin{align}
\s = u \bpmat 1+\frac{n}{2\pi i}
\ln\left(\frac{u}{\L}\right)\\ 1 \epmat.
\end{align}
For this solution the metric is $ds^2 = -\frac{n}{4\pi}\left\{\text{ln}\left(\frac{u\ub}{\Lambda^2}\right) +2\right\}dud\ub$.  Note that as $|u|\to 0$, ln$(u\ub) \to -\infty$, so the metric is positive-definite in the vicinity of $u=0$ only for $n>0$.  Thus the $T^n$ monodromies for $n \in \Z^+$ give sensible geometries.  In this case $\L$ is the Landau pole.  A similar story goes for the $-T^n$ monodromies.  They give positive definite metrics for $n\in \Z^+$, and which correspond, as we saw in exercise \ref{MonSu2}, to $\suf(2)$ theories with $n+4$ massless fundamental hypermultiplets.

\begin{ex}[]{}
\begin{exe}\label{SL(2,Z)}
Eigenvalues of $SL(2,\Z)$ matrices always come in pair $(\lambda_1,\lambda_2)$ such that $\lambda_1 \lambda_2=1$ and $SL(2,\Z)$ elements can be consequently characterized as:
\begin{enumerate}[leftmargin=3cm,rightmargin=.3cm]
\item[\textsc{Elliptic}] If $\lambda_{1,2}\in\C$ then $\lambda_2=\lambda_1^*=\lambda$ and $|\lambda|=1$.

\item[\textsc{Hyperbolic}] If $\lambda_{1,2}\in \R$ then $\lambda_2=1/\lambda_1=r$.

\item[\textsc{Parabolic}] If $\lambda_1=\lambda_2=\pm1$ and the matrix can only be reduced to a Jordan form.
\end{enumerate}

Compute explicitly that the allowed eigenvalues for an elliptic element of $SL(2,\Z)$ are:
\beq
\l=\pm1,e^{\pm i2\pi/3},e^{\pm i\pi/2},e^{\pm i\pi/3}.
\eeq
A similar characterization of its elements extend to $\Sp(2r,\Z)$ and its elleptic elements also have their eigenvalues strongly constrained which similarly constrains the set of scaling dimensions of CB coordinates at arbitrary ranks \cite{Argyres:2018urp,Caorsi:2018zsq}.
\end{exe}

\begin{center}
\rule[1mm]{2cm}{.4pt}\hspace{1cm}$\circ$\hspace{1cm} \rule[1mm]{2cm}{.4pt}
\end{center}

\begin{exe}\label{SWcurve}
Come up with a physical argument to convince yourself that for a rank-1 $\cN=2$ SCFTs $\tau$ should be constant on the whole CB. Does this argument also work for rank-2?  (Hint: think about RG-flows)
\end{exe}

\begin{center}
\rule[1mm]{2cm}{.4pt}\hspace{1cm}$\circ$\hspace{1cm} \rule[1mm]{2cm}{.4pt}
\end{center}

\begin{exe}\label{KahMet}
Using the globally defined CB coordinate $u$, compute the K\"ahler metric associated to the various $\cN=2$ SCFTs that we studied and show that, despite $\tau$ being constant, the metric has a singularity at the origin and one at infinity. What is wrong with computing the metric using the special coordinates?
\end{exe}
\end{ex}


\section{Seiberg-Witten curve and higher ranks}
\setcounter{exe}{0}

We are now ready to finally introduce the Seiberg-Witten (SW) curve and the SW one-form, outline the generalization of special K\"ahler geometry for higher ranks as well as make connection with the construction of the CB of class-S theories. It is important to be clear in that Special K\"ahler geometry captures the full scope of the constraints of low-energy $\cN=2$ SUSY, the class-S and SW story should be understood as (some times very effecient) \emph{tools} to explicitly construct the Special K\"ahler geometries in examples. In this section we will assume some knowledge of basic algebraic geometry, good introductory texts are \cite{Griff:1978,Shafa:1977}.

\subsection{SW curve and SW one-form}

\begin{figure}
\begin{center}
\includegraphics[width=.4\textwidth]{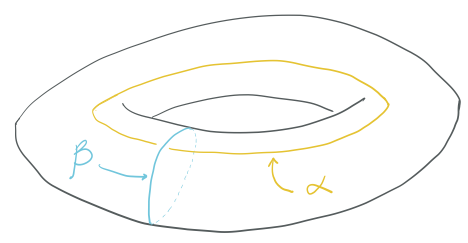}
\caption{\label{cycles} Depiction of the $\a$ and $\b$ cycles of $T^2$.}
\label{perio}
\end{center}
\end{figure}

An important observation made in \cite{Seiberg:1994aj,Seiberg:1994pq} is that there are two properties of a rank-1 torus which very closely resemble special K\"ahler geometry:
\begin{itemize}
\item[1)] There is a one-to-one correspondence between $\SL(2,\Z)$ equivalence classes of complex numbers with positive imaginary part and complex structures of rank-1 tori, $T^2$. The $\SL(2,\Z)$ action on $\tau\in\C$ is precisely the fractional linear transformation \eqref{tauSLtwoZ}:
\beq\label{prop1}
\tau\cong \tau'=\frac{m_1 \tau+m_2}{m_3 \tau+m_4}\quad {\rm with} \quad m_1m_4-m_2m_3=1.
\eeq

\item[2)] The periods of the torus, which are defined as the integration of the holomorphic one-form on $T^2$ over the two one-cycles generating the first homology (see figure \ref{cycles}):
\beq
b_1=\oint_\alpha \l_{\rm hol}\qquad\&\qquad b_2=\oint_\beta \l_{\rm hol}
\eeq
satisfy the following properties:
\beq\label{prop2}
\frac{b_1}{b_2}=\tau\quad\&\quad \tau\to\tau'\quad\Rightarrow\quad \left(
\begin{array}{c}
b_1\\
b_2
\end{array}
\right)\to
\left(
\begin{array}{cc}
m_1&m_2\\
m_3&m_4
\end{array}
\right)
\left(
\begin{array}{c}
b_1\\
b_2
\end{array}
\right)
\eeq
\end{itemize}
We used the letter $\t$ to indicate the complex structure of the torus not coincidentally. The reader might have already noticed the similarities between the above properties and those of the holomorphic gauge coupling defining the K\"ahler metric on $\cC$.

\eqref{prop1} and \eqref{prop2} are very resembling of the properties defining special K\"ahler geometry and they almost make it for a direct identification between $a^D$ and $a$ and the periods. To make the correct identification, notice that nowhere in discussing the torus the CB parameter appeared. To fully identify the complex structure of a torus with the low-energy holomorphic gauge coupling on $\cC$  we need to somehow obtain a non-trivial $u$ dependence of the former. This is easily achieved by considering a non-trivial fibration of tori over $\cC$, with possibly varying complex structure, and we call $\tau(u)$ the complex structure of the torus fibered over the point $u$. Similarly $b_1(u)$ and $b_2(u)$ will be the periods computed for the torus at $u$.

\begin{figure}
\begin{center}
\includegraphics[width=.9\textwidth]{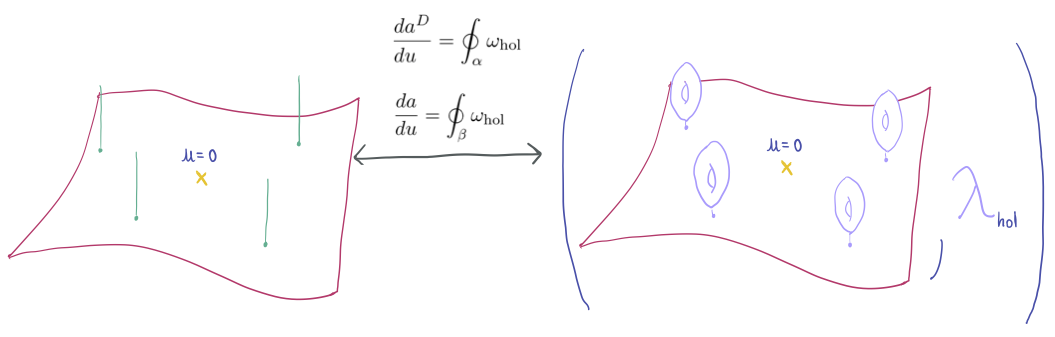}
\caption{\label{Map_bund} The holomorphic section of the $\SL(2,\Z)$ vector bundle characterizing the special K\"ahler geometry can be constructed by fibering a $T^2$ at each point and computing its periods.}
\label{Mapb}
\end{center}
\end{figure}

Secondly notice that:
\beq
\tau(u)=\frac{d a^D}{da\phantom{^D}}=\frac{d a^D}{du\phantom{^D}}\frac{d u}{da}
\eeq
this simple observation, makes the following identifications obvious:
\beq\label{PerInt}
\frac{d a^D}{du\phantom{^D}}=\oint_\alpha \w_{\rm hol}\qquad\&\qquad  \frac{d a}{du}=\oint_\beta \w_{\rm hol}
\eeq
arriving at the final picture which is summarized in figure \ref{Map_bund}. The Special K\"ahler geometry can be reconstructed by a pair $(\Ssw,\lsw)$, the \emph{SW curve} and \emph{SW one-form}. The former being a fibration of tori over the CB $\cC$ (more below) and the latter being a one form satisfying\footnote{Given what we discussed thus far, it might seem a bit artificial to make the distinction between $\lsw$ and $\w_{\rm hol}$. This is because we are considering the conformal case in which this $\lsw$ is also holomorphic. The situation changes in the presence of masses. In that case $\lsw$ is in fact not holomorphic and \eqref{swc} is a non-trivial relation.}
\beq\label{swc}
\frac{d\lsw}{du}=\w_{\rm hol}\quad\Rightarrow
\left\{
\begin{array}{l}
a^D=\oint_\alpha \l_{\rm SW}\\
a=\oint_\beta \l_{\rm SW}
\end{array}
\right.
\eeq
For the rank-1 case which we have so far discussed, this perspective is particularly useful. There is in fact a closed algebraic form for $\S$. This works as follows.

Genus-1 tori are \emph{elliptic curves}, which can be embedded in $\C^2$ as follows:\footnote{To avoid discussing complex projective spaces we allowed ourselves to be a bit sloppy. Elliptic curves are in fact defined in $\C P^2$ and are therefore written as zeros of a single polynomial in three variables defined up to a overall multiplication of a non-zero complex number. The form in \eqref{Wei}, which strictly speaking only extends over a patch of $\C P^2$, is almost correct but it misses \emph{the point at infinity}. }
\beq\label{Wei}
y^2=x^3+\#_1 x+\#_2
\eeq
$\#_1$ and $\#_2$ are complex coefficients which determine the complex structure of the torus $\tau$. \eqref{Wei} is called the Weierstrass form of the elliptic curve. 

As we discussed, the total space $\Ssw$ is obtained by fibering a $T^2$ over any point of $\cC$. This can be readily obtained by making $\#_1$ and $\#_2$ functions of the CB parameter $u$. Because of the connection between genus-1 tori and elliptic curves, fibering a genus-1 torus over a space is often referred to as \emph{ellipitc fibration}. Thus the most generic $\Ssw$ has the form:
\beq\label{SWc}
\Ssw:\quad y^2=x^3+f(u)x+g(u)
\eeq
where $f$ and $g$ are holomorphic functions of $u$. 

Since on $T^2$ there is only one holomorphic one form, the SW one form is easily written down:
\beq\label{SWof1}
\w_{\rm hol}=\frac{dx}{y}=\frac{dx}{\sqrt{x^3+f(u) x+g(u)}}=\frac{d \lsw}{du}.
\eeq
To get some familiarity with the expression \eqref{SWc} and \eqref{SWof1}, it is useful to work out the curves for the rank-1 geometries in table \ref{Kodaira}, which match the entries in table \ref{table:eps}, see exercise \ref{SWcurves}. This can be done extending the $\C^*$ action \eqref{Csrank1} to the total space $\Ssw$ and applying dimensional analysis.


\begin{table}
\centering
\def\arraystretch{1.2}
$\begin{array}{|c|c|c|l|}
\hline
\ \D_u\ \,& \mathscr{M}_\g               & \t         & \multicolumn{1}{c|}{{\rm SW curve}} \\
\hline
6           & ST              & e^{i\pi/3} &\ y^2=x^3+u^5    \\
4           & S                & i              & \    y^2=x^3+u^3 x  \\
3           & (-ST)^{-1} & e^{2i\pi/3} &\ y^2=x^3+u^4      \\
2           & -\I               & \ \text{any}\ \t\ \,&\ y^2 \prod_{i=1}^3\left(x-e_i(\t)\, u\right)  \\
3/2        & -ST            & e^{2i\pi/3}   & \  y^2=x^3+u^2  \\
4/3        & S^{-1}      & i                  &\  y^2=x^3+u x   \\
6/5        & (ST)^{-1}  & e^{i\pi/3}   &\ y^2=x^3+u  \\
1           & \I                & \text{any}\ \t &\ y^2=x^3+u  \\
\hline
\end{array}$
\caption{Possible values of $\D_u$, $M$, $\t$, $a^D$ and $a$ for rank-1 CB singularities.\label{Kodaira} Below the dashed line are the IR-free geometries which are discussed below.}
\end{table}

\paragraph{Performing the period integrals.} Before concluding this subsection, it is useful to explain how to compute the integrals in \eqref{PerInt} and obtain an expression for the special coordinates given $(\Ssw,\lsw)$. First re-write \eqref{Wei} as
\beq
y^2=(x-e_1)(x-e_2)(x-e_3)
\eeq
a bit of complex geometry reveals that an elliptic curve is a double cover of the Riemann sphere (that is $\C$ with the point at infinity) branched over four points ($e_1,e_2,e_3,\infty$). Then the $\a$ and $\b$ cycles in figure \ref{cycles} can be identified as the two loops in figure \ref{PerInt} and the two period integrals become:
\begin{align}\label{Adcomp}
\frac{da^D}{du\phantom{^D}}=\oint_\a \frac{dx}{y}&=\frac{4}{\sqrt{e_1-e_3}}K(k_\a)\\\label{Acomp}
\frac{da}{du}=\oint_\b \frac{dx}{y}&=\frac{4}{\sqrt{e_3-e_1}}K(k_\b)
\end{align}
where the elliptic integral $K(k)$ is
\beq\label{ElliInt}
K(k)=\int^1_0\frac{dx}{\sqrt{(1-x^2)^2(1-k^2x^2)}},
\eeq
and
\begin{align}
k_\a^2&=\frac{e_2-e_3}{e_1-e_3}\\
k_\b^2&=\frac{e_2-e_1}{e_3-e_1}
\end{align}

\begin{figure}
\begin{center}
\includegraphics[width=.6\textwidth]{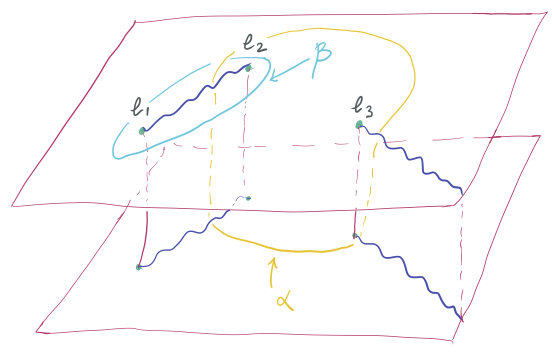}
\caption{\label{PerInt} Depiction of the one-cycles on the branched double cover of the Riemann sphere.}
\end{center}
\end{figure}

To obtain numerical results, the integral \eqref{ElliInt} is tabulated in Mathematica as $K(k)=\texttt{EllipticK[}k^2\texttt{]}$.


\subsection{Generalization to higher ranks}

Let us now outline briefly how the CB geometry works if dim$_{\C}\cC=r\neq1$. As we discussed at the end section \ref{sec:modsp}, at sufficient low-energy, the effective theory on a generic point of the CB is a $\cN=2$ $U(1)^r$ theory, where $r$, the \emph{rank} of the theory, is precisely the complex dimension of the CB. This effective theory is described by $r$ $\cN=2$ vector multiplets. The K\"ahler potential of the low energy effective theory again defines a K\"ahler metric on $\cC$ which acquires the form (see \eqref{metrRr}):
\beq
ds^2={\rm Im} \tau_{IJ}d\bar{a}^I\otimes da^J,\quad I,J=1,...,r.
\eeq 
where $a_I$ are identified as vevs of the scalar component of the $r$ $\cN=2$ vector multiplets describing the $r$ massless photons in the low-energy limit\footnote{Notice that the $\cN=2$ vector multiplet of a $U(1)$ theory is in fact gauge invariant, thus $a$ is a legitimate observables.} and $\tau_{IJ}$ are physically interpreted as the holomorphic gauge couplings of the $r$ $U(1)$ factors:
\beq
\t_{IJ}=\frac{i4\pi}{g^2_{IJ}}+\frac{\theta_{IJ}}{2\pi}\qquad
{\rm where}\qquad\begin{array}{l}
i)\ {\rm Im}\t_{IJ}>0.\\[5pt]
ii)\ \t_{IJ}=\t_{JI}
\end{array}
\eeq
Condition $ii$) is trivially satisfied at rank-1 and is obvious from the physics but it adds a non trivial condition on the geometry at rank higher than 1, see for instance \cite{Argyres:2005pp,Argyres:2005wx}. The geometry of the CB is again captured more directly introducing the special coordinates and in particular a set of $r$ dual coordinates $a^D_I$, $I=1,...,r$ which satisfy:
\beq
\t_{IJ}=\frac{\partial a^D_I}{\partial a^J}
\eeq
The electromagnetic duality group generalizes to $Sp(2r,\Z)$, see section \ref{EMduality} where a convenient way of parametrize this group was presented - \eqref{Sp2rZ1} and \eqref{Sp2rZ2}. Using ${\boldsymbol a}^D$ and ${\boldsymbol a}$ to indicate an $r$-component vector and ${\boldsymbol \tau}=\partial {\boldsymbol a}^D/\partial {\boldsymbol a}$, $\Sp(2r,\Z)$ acts on $\t$ and the special coordinates as follows:
\beq
\left(\begin{array}{c}
{\boldsymbol a}^D\\
{\boldsymbol a}
\end{array}\right)\to\left(
\begin{array}{cc}
A&B\\
C&D
\end{array}
\right)
\left(\begin{array}{c}
{\boldsymbol a}^D\\
{\boldsymbol a}
\end{array}\right),\qquad
\&\qquad
{\boldsymbol\tau}\rightarrow{\frac{A{\boldsymbol \tau} + B}{C{\boldsymbol \tau} + D}}\,,
\label{tauSp2rZ}
\eeq

This structure is encoded with the arbitrary rank generalization of Definition \ref{SpecialKa}, which goes as follows.

\begin{definition}[Special K\"ahler space - rank-r]{SpecialKaRar}
We define a Singular Special K\"ahler space as a triple $(\cC,\sigma,\tau)$ where:\
\begin{itemize}
\item $\cC$ is a K\"ahler space of complex dimension $r$ with globally defined coordinates $\boldsymbol{u}=(u_1,...,u_r)$.

\item ${\boldsymbol \sigma}(\boldsymbol{u}):=\left(\begin{array}{c}{\boldsymbol a}^D(\boldsymbol{u})\\{\boldsymbol a}(\boldsymbol{u})\end{array}\right)$ is a 2$r$ dimensional holomorphic section of an $\Sp(2r,\Z)$ bundle over $\cC$.

\item $\bt:=\frac{\partial \bsa^D}{\partial \bsa}$ is a $r\times r$ matrix such that:
\begingroup
\setlength\abovedisplayskip{0pt}
\begin{align}\nonumber
i)&\ {\rm Im}\ (\bt)>0. \\\nonumber
ii)&\ \t_{IJ}=\t_{JI}.\\\nonumber
iii)&\ ds^2=({\rm Im}\bt) d\bar{\bsa}\otimes d\bsa.
\end{align}
\endgroup
\end{itemize}
\end{definition}

\begin{figure}
\begin{center}
\includegraphics[width=.8\textwidth]{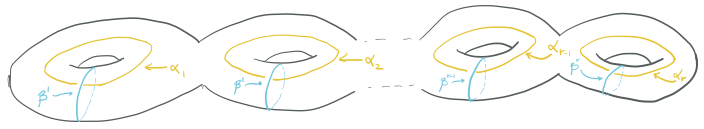}
\caption{$H_1(\Csc_r,\Z)\equiv \Z^{2r}$ and we can choose a base of the first homology such that $\a_I\cdot\b^J=\d^J_I$ and $\a_I\cdot\a_J=\b^I\cdot\b^J=0$}
\label{genusr}
\end{center}
\end{figure}

It is natural for the reader to wonder whether there is a generalization of the SW curve and one-form story to the theories of general rank. The answer is yes and it works as follows. The elliptic fibration of rank-1, see figure \ref{Map_bund}, is generalized to a fibration of a genus-$r$ Riemann surface $\Csc_r$ on a $r$-complex dimensional base which will be identified with the CB $\cC$ of the theory, see figure \ref{genusr}. In this case $H_1(\Csc,\Z)\equiv \Z^{2r}$ so we can choose a basis of 1-cycles $(\a_I,\b_I)$ $I=1,...,r$ whose intersections can be appropriately normalized: $\a_I\cdot \b^J=\delta^J_I$, $\a_I\cdot\a_J=\b^I\cdot\b^J=0$. $\Csc_{r}$ has genus-$r$ and it admits $r$ independent holomorphic one-forms $\w_{\rm hol}^I$.\footnote{The fact that a genus $r$ compact Riemann surface $\Csc_r$ has precisely $r$ independent holomorphic one-forms can be seen, for instance, applying the Riemann-Roch theorem to the case of the canonical bundle $K\to \Csc_r$. This fact gives an algebraic definition of the genus. Deriving this result is relatively straightforward and it is useful to start appreciating the power of algebraic geometry.} Then periods are straightforwardly generalized to the following
\beq
b_I=\oint_{\a_I} \w_{\rm hol}^I,\qquad b^I=\oint_{\b^I} \w_{\rm hol}^I,\qquad I,=1,...,r.
\eeq
and all the relevant properties also follow. We could thus define the SW one-form
\beq
\l_{\rm SW}:\quad \frac{\partial \l_{\rm SW}}{\partial u_I}=\w^I
\eeq
which leads to a straightforward generalization of \eqref{swc}:
\beq
a^D_I:=\oint_{\a_I}\l_{\rm SW}\quad \&\quad a^I:=\oint_{\b^I}\l_{\rm SW}
\eeq
correctly reconstructing the SW geometry for higher rank theories \cite{Klemm:1994qs,Argyres:1994xh,Argyres:1995fw,Argyres:1995wt,Danielsson:1995is,Hanany:1995na,Brandhuber:1995zp,Donagi:1995cf}.

A few remarks are in order. First, in this case the SW curve is considerably more complicated. In fact, there is no simple close algebraic description of a generic fibration of a compact genus $r$ Riemann surface $\Csc_r$ over a $r$ complex dimensional dimensional base $\cC$ for $r>2$. Since all genus 2 curves are given by a single equation of the form $y^2=\mathscr{P}(x)$, where $\mathscr{P}(x)$ is a polynomial of degree at most six\footnote{Algebraic curves which can be written in this form are called \emph{hyperelliptic} if the degree of $\mathscr{P}(x)$ is strictly greater than four.},  at rank-2 such a fibration can be written as a two complex dimensional family of polynomial equations. 

Secondly in describing the generalization above, we have been a bit quick. In fact the type of fibration that we described involving compact Riemann surfaces $\Csc_r$ of genus $r$, only captures a subset of measure zero of possible Special K\"ahler geometry. The $\tau_{IJ}$ satisfying the condition in Definition \ref{SpecialKaRar}, is in one to one correspondence to a close cousin of complex structures compact Riemann surface of genus $r$; rank-$r$ polarized abelian variety. The correct generalization of the SW story to rank-$r$ is a fibration of rank-$r$ polarized abelian varieties rather than genus $r$ Riemann surfaces and this difference starts becoming important for $r>3$. These objects are just rank-$r$ algebraic complex tori\footnote{A rank-$r$ complex torus is defined in a similar way the more familiar $T^2$ as $\C^r/\L$, where $\L$ is a rank-$2r$ lattice. Complex tori are complex manifold but they cannot all be written as a space of solutions of a set of algebraic equations. Those which can, are algebraic varieties.} and in many cases are related to $\Csc_r$, but rank-$r$ polarized abelian varieties represent in fact a far larger set than $\Csc_r$.

Without getting too deep in algebraic geometry technicalities, let us elaborate a bit more on this point. A first helpful observation is that there is a canonical map which associate a rank-$r$ polarized abelian variety to a genus $r$ compact Riemann surface. This map is called the Abel-Jacobi map and the resulting abelian variety is called the \emph{Jacobian variety} of $\Csc_r$ and indicated by $J(\Csc_r)$. The statement that the set of rank-$r$ polarized abelian varieties is larger than that of genus-$r$ compact Riemann surfaces, is the statement that the Abel-Jacobi map is not surjective and the set of Jacobian varieties $J(\Csc_r)$ is a set of measure zero in the set of rank-$r$ abelian varieties. 

A second important observation is that if we replace the $\Csc_r$ fiber by a rank-$r$ abelian variety, something more radical has to happen in the way we define both the K\"ahler metric and the special coordinates. In fact rank-$r$ abelian varieties are $r$ complex dimensional spaces and it is no longer the case that we can integrate a one-form over it. The framework to properly understand this generalization is that of \emph{complex integrable systems} \cite{Donagi:1995cf}. Roughly speaking a generalization of action-angle coordinates to the complex set-up. For lack of time, we will not delve in describing the connection between integrable systems and CB geometry. This perspective will return to bite us again in the next subsection where we will summarize the way in which the CB geometry for a large set of $\cN=2$ SCFTs, so-called \emph{theories of class-}$S$ \cite{Gaiotto:2009we,Gaiotto:2009hg}, can be effectively computed.The peculiar feature of theories of class-$S$ is that they are obtained from compactification of six dimensional $(2,0)$ theories. 


\subsection{Class-S theories}

Here we will review how to construct the CB for class-S. Recall that the class-S construction \cite{Gaiotto:2009we,Gaiotto:2009hg} involves starting with a $\gf$ (2,0) theory in six dimension, putting it on $\R^{3,1}\times\Csc$, where $\Csc$ is a Riemann surface. The $\cN=2$ theory of our interest is the theory obtained in the limit in which the volume of $\Csc$ shrinks to zero. The important properties of the six dimensional (2,0) theory to understand the brief description below are:

\begin{tcolorbox}
\begin{center}
\textbf{Properties of 6d (2,0) theories}
\end{center}\vspace{-.5cm}
\begin{itemize}[leftmargin=*]
\item[\textbf{\dred{a.}}] (2,0) theories are maximally supersymmetric theories in six dimensions and are identified by a simply-laced lie algebra $\gf$ and will be labeled as $\cT[\gf]$, $\gf=\{A_n,D_n,E_6,E_7,E_8\}$.

\item[\textbf{\dred{b.}}] $\cT[\gf]$ is superconformal and the (2,0) superconformal algebra in six-dimensions is $\mathfrak{osp}(8|4)$ with $R$-symmetry $\sof(5)_R$.

\item[\textbf{\dred{c.}}] $\cT[\gf]$  has a $5 r$ real dimensional moduli space, with $r$ being the rank of $\gf$. 

\item[\textbf{\dred{d.}}] The low-energy theory on this moduli space is that of $r$ free tensor multiplet. For this reason this is often call a \emph{tensor branch}.

\item[\textbf{\dred{e.}}] The free tensor multiplet contains a scalar component transforming the in the ${\bf 5}$ of $\sof(5)_R$, we will call that $\varphi$.

\item[\textbf{\dred{f.}}] Upon circle compactification the $\cT[\gf]$ flows to a maximally supersymmetric $\gf$ Yang-Mills theory on $\R^{4,1}$.
\end{itemize}
\end{tcolorbox}\vspace{0.5em}

The moduli space of $\cTg$ has a simple description:
\beq\label{tensBra}
\cM_{\rm tensor}=\R^{5r}/{\rm Weyl}(\gf)
\eeq
where $r$ is the rank of the (simply laced) Lie algebra $\gf$ characterizing the six dimensional theory. The factor $\R^5$ can be understood by the fact that the scalar component of the tensor multiplet, as we said above, transforms in the fundamental of the $\sof(5)_R$ R-symmetry. As we saw in the case of the CB of $\cN=2$ theories, the space in \eqref{tensBra} is parametrized by the generators of the invariant polynomials in $\R^{5r}$ under Weyl$(\gf)$. Those are, roughly speaking, the vevs of the gauge invariant operators 
\beq\label{Oper}
\cO_d:={\rm tr}\big(\varphi^d\big)
\eeq 
for appropriate values of $d$ (we are here suppressing an $\sof(5)_R$ index). 

To reduce to four dimension we need to put the theory $\R^{5,1}\to\R^{3,1}\times\Csc$ where $\Csc$ is a compact Riemann surface whose topology is specified by two integers $(g,p)$, the former indicating the genus of the $\Csc$ and the latter the number of punctures (we won't discuss at all how $g$ and $p$ characterize the final four dimensional theory). Right away the initial $\sof(5,1)$ Poincar\'e symmetry of the original six dimensional theory is broken down to $\sof(3,1)\oplus \sof(2)_{\Csc}$. Similarly the $R$-symmetry of the $(2,0)$ theory decomposes $\sof(5)_R\to\sof(3)_R\oplus\sof(2)_R$, which gives the correct $R$ symmetry for a four dimensional $\cN=2$ theory. The scalar of the tensor multiplet, splits as ${\bf 5}\to {\bf 3}\oplus {\bf 2}$ where the ${\bf 2}$ can be rearranged in a single complex scalar which we call $\phi$. Projecting \eqref{Oper} down to the ${\bf 2}$ component, we get the operators whose vevs parametrize the CB of the corresponding four dimensional theory (which are then some homogenous polynomials in $\phi$ and $\bar{\phi}$). 

Things are a bit subtler. The curvature of the Riemann surface breakes supersymmetry (think about why?). To land onto a $\cN=2$ theory in four dimension, we then need to implement a partial twist \cite{Witten:1988ze}. This involves identifying the $\sof(2)_\Csc$ with the $\sof(2)$ component of the $R$-symmetry. The consequence of this twisting is that all local operators carrying an $\sof(2)_R$ charge also acquire non-trivial transformation under the diffeomorphism of $\Csc$. In particular a local operator carrying charge $r$ under $\sof(2)_R$, transforms now as a $r$ differential on $\Csc$. One convenient way to account for the partial twisting is to elevate the operators in \eqref{Oper} which parametrize the CB, to degree $d$ differential on $\Csc$. 

The partial twisting condition also sets to zero all non-holomorphic vevs and in particular the initial scalars of the 6d theories which could acquire vevs are now $\gf$-valued holomorphic one form on $\Csc$
\beq\label{redu}
\varphi\quad\to\quad \phi(z)dz
\eeq
where we have suppressed both the Lie algebra and the $R$-symmetry indices. We are then left with the fact that the CB of the resulting four dimensional theory is parametrized by the vevs of holomorphic $d$ differentials on $\Csc$. These objects are counted by $H^0(\Csc,K^{\otimes^d})$, which formally is the 0-th cohomology group of the sheaves of section of various tensor product of the canonical bundle over $\Csc$ and precisely counts the global holomorphic $d$-differential over $\Csc$. Thus the CB of class-S is often written, in a somewhat fancy form, as:
\beq\label{CBclaS}
\cB=\bigoplus_iH^0(\Csc,K^{\otimes^{d_i}})
\eeq
and where the $d_i$s are precisely given by the degrees of the polynomial invariant under Weyl$(\gf)$ in \eqref{Oper}. Notice that we should have used $\cC$ but we chose $\cB$ instead to make connection with class-S notation.

\eqref{CBclaS} gives the parametrization of the space of vacua of the theory. But how do we reconstruct the full special K\"ahler geometry? Ultimately, if \eqref{CBclaS} is truly to be interpreted as a CB of a $\cN=2$ theory, it should also carry the extra information of the low-energy $U(1)^r$ effective theory which is encoded in the special coordinates. To really understand this point, there is a lot of extra background needed which we won't be able to cover. I will instead sketch the mean points, unfortunately losing in pedagogy.

As mentioned in passing in the previous section, the SW geometry in general defines a complex integrable systems. This object is readily accessible in the class-S description by an elaborate set of non-trivial observations involving 5d, 4d and 3d supersymmetric theories. The gist of it is that there is an advantage of studying the theory obtained by compactifying the 4d theory on a circle. This is because we can ``invert'' the order of compactification, that is put the initial (2,0) theory on $\R^{2,1}\times\Csc\times S^1$ and first shrink the $S^1$, and take advantage that the theory we obtain is a five dimensional $\gf$ Yang-Mills theory on $\R^{2,1}\times\Csc$ which has a lagrangian description and it is easier to study. By studying a 5d $\gf$ Yang-Mills theory on $\R^{2,1}\times \Csc$, it is possible to precisely derive the CB of the theory in three dimensions, which is $\cN=4$ supersymmetric\footnote{A quick way to understand the amount of supersymmetry preserved in three dimension, is to notice that we can get there by compactifying the 4d theory on a circle. This compactification does not break any supersymmetry and therefore the 3d theory should contain the same amount of supersymmetry as an $\cN=2$ theory. Since $\cN=n$ in 4d corresponds to $\cN=2n$ in 3d, this readily gives the answer.}. We call this space $\cM_{3d,\Csc}$ which, by $\cN=4$ supersymmetry in 3d, is an hyperk\"ahler space.

There is a canonical way to reconstruct the complex integrable system arising from the CB of the 4d theory from the hyperk\"ahler CB of the 3d theory\footnote{The details are inessentials but for the curious reader this involves identifying a distinguished complex structure carried by $\cM_{3d,\Csc}$ and we reconstruct the complex integrable system by considering $\cM_{3d,\Csc}$ as a holomorphic symplectic variety with respect to this complex structure}. Therefore the knowledge of $\cM_{3d,\Csc}$ allows to reconstruct the full SW geometry of the four dimensional theory.

A bit more explicitly, the complex integrable system that we obtain in the class-S construction, is of special type, namely a family of Hithcin systems \cite{Hitchin:1987mz}. These are very well studied in mathematics and are precisely obtained fibering a polarized abelian variety over the affine space given in \eqref{CBclaS}. We can be even more explicit in what the abelian variety looks like, it is in fact the Jacobian variety associated to a Riemann surface which isn't quite $\Csc$, rather a (ramified) covering $\Sigma\to \Csc$ of it\footnote{Again, this statement is not quite precise. The covering $\Sigma\to\Csc$ induces a map between the two associated Jacobian varieties $J(\Sigma)\to J(\Csc)$. The fiber of the fibration is the subvariety of $J(\Sigma)$ which is the kernel of this map or also called the \emph{Prym variety} of the covering.}. This ramified covering can be written algebraically as the \emph{spectral curve} (of the Higgs bundle associated to the Hitchin system):
\beq
{\rm det}(\phi(z)dz-xdz)=0
\eeq 
where $\phi(z)$ is the $\gf$-valued holomorphic one-form in \eqref{redu}. Notice that $x$ parametrizes a generic component of a one-form on $\Csc$. The covering $\Sigma\subset T^*\Csc$, the cotangent bundle of the Riemann surface $\Csc$. 

Finally the SW one-form can be obtained from the symplectic two form which is canonically given by the family of integrable systems which in this case is the family of Hitchin systems. Doing this procedure carefully we obtained that:
\beq
\l_{\rm SW} \equiv x dz
\eeq
This very abstract presentation can be made concrete by choosing specific examples and, for those theories of class-S for which a lagrangian theory is known, it has been in fact checked that this picture actually reproduces the SW geometry of the previous sections.

\begin{ex}[]{}
\begin{exe}\label{SWcurves}
Derive the SW curves in table \ref{Kodaira} for all the entries in table \ref{table:eps}. This can be done using straightforward scaling argument. First extend the scaling action \eqref{Csrank1} on $(u,a,a^D)$ to the the coordinates $(x,y)$ of the family of elliptic curves and show that this $\C^*$ action on the total space $\Ssw$ is given by:
\begin{align}\label{cond1}
2[y]&=3[x]=[f(u)]+[x]=[g(u)]\\\label{cond2}
f(u)=&u^m\quad \&\quad g(u)=u^\ell,\qquad m,\ell \in \N
\end{align}
where $[\cdot]$ indicates weight of $\cdot$ under the $\C^*$ action. Then show that \eqref{cond1} and \eqref{cond2} imply that $m$ and $\ell$ cannot be non-zero at the same time, unless $[u]=2$ (which already singles out this case). Finally, imposing (why is it that $[a]=[a^D]=1$?):
\beq
[\w_{\rm hol}]=1-[u],\qquad [u]\geq1
\eeq
and using \eqref{SWof1}, derive the curves in table \ref{Kodaira} as only allowed possibilities.
\end{exe}

\begin{center}
\rule[1mm]{2cm}{.4pt}\hspace{1cm}$\circ$\hspace{1cm} \rule[1mm]{2cm}{.4pt}
\end{center}

\begin{exe}\label{SWcurve}
Compute explicitly $(a^D,a)$ using the expression for the curves in table \ref{Kodaira} and \eqref{Adcomp} and \eqref{Acomp}. Show that, up to an inessential $u$ independent numerical coefficient, the expression for the special coordinates coincide what we derived previously in table \ref{table:eps}.
\end{exe}
\end{ex}

\acknowledgments

The author would like to thank the organizers and the students of the \emph{Young Researchers Integrability School (YRISW 2020): A modern primer for superconformal field theories} for the wonderful time spent there despite the cold and rainy Hamburg weather. I also would like to thank Grant Elliot for comments on the lectures notes. It is finally an absolute pleasure to thank Philip Argyres for over five years and counting of a productive collaboration and fulfilling friendship. My perspective on the subject is certainly most influenced by Philip's brilliant insights. MM is supported by NSF grants PHY-1151392 and PHY1620610.

\appendix

\bibliographystyle{JHEP}

\begin{thebibliography}{10}

\bibitem{Razamat:2020pra}
S.~S. Razamat, E.~Sabag, and G.~Zafrir, {\it {Weakly coupled conformal
  manifolds in 4d}},  \href{http://arxiv.org/abs/2004.07097}{{\tt
  arXiv:2004.07097}}.

\bibitem{Argyres:1995xn}
P.~C. Argyres, M.~R. Plesser, N.~Seiberg, and E.~Witten, {\it {New N=2
  superconformal field theories in four-dimensions}},  {\em Nucl. Phys.} {\bf
  B461} (1996) 71--84, [\href{http://arxiv.org/abs/hep-th/9511154}{{\tt
  hep-th/9511154}}].

\bibitem{Argyres:1995jj}
P.~C. Argyres and M.~R. Douglas, {\it {New phenomena in SU(3) supersymmetric
  gauge theory}},  {\em Nucl. Phys.} {\bf B448} (1995) 93--126,
  [\href{http://arxiv.org/abs/hep-th/9505062}{{\tt hep-th/9505062}}].

\bibitem{Gaiotto:2009we}
D.~Gaiotto, {\it {N=2 dualities}},  {\em JHEP} {\bf 1208} (2012) 034,
  [\href{http://arxiv.org/abs/0904.2715}{{\tt arXiv:0904.2715}}].

\bibitem{Gaiotto:2009hg}
D.~Gaiotto, G.~W. Moore, and A.~Neitzke, {\it {Wall-crossing, Hitchin Systems,
  and the WKB Approximation}},  \href{http://arxiv.org/abs/0907.3987}{{\tt
  arXiv:0907.3987}}.

\bibitem{Garcia-Etxebarria:2015wns}
I.~Garc{\'i}a-Etxebarria and D.~Regalado, {\it {$ \mathcal{N}=3 $ four
  dimensional field theories}},  {\em JHEP} {\bf 03} (2016) 083,
  [\href{http://arxiv.org/abs/1512.06434}{{\tt arXiv:1512.06434}}].

\bibitem{Aharony:2015oyb}
O.~Aharony and M.~Evtikhiev, {\it {On four dimensional N = 3 superconformal
  theories}},  {\em JHEP} {\bf 04} (2016) 040,
  [\href{http://arxiv.org/abs/1512.03524}{{\tt arXiv:1512.03524}}].

\bibitem{Argyres:2016xua}
P.~C. Argyres, M.~Lotito, Y.~L{\"u}, and M.~Martone, {\it {Expanding the
  landscape of $ \mathcal{N} $ = 2 rank 1 SCFTs}},  {\em JHEP} {\bf 05} (2016)
  088, [\href{http://arxiv.org/abs/1602.02764}{{\tt arXiv:1602.02764}}].

\bibitem{Aharony:2016kai}
O.~Aharony and Y.~Tachikawa, {\it {S-folds and 4d N=3 superconformal field
  theories}},  {\em JHEP} {\bf 06} (2016) 044,
  [\href{http://arxiv.org/abs/1602.08638}{{\tt arXiv:1602.08638}}].

\bibitem{Bonetti:2018fqz}
F.~Bonetti, C.~Meneghelli, and L.~Rastelli, {\it {VOAs labelled by complex
  reflection groups and 4d SCFTs}},
  \href{http://arxiv.org/abs/1810.03612}{{\tt arXiv:1810.03612}}.

\bibitem{Argyres:2019ngz}
P.~C. Argyres, A.~Bourget, and M.~Martone, {\it {Classification of all
  $\mathcal{N}\geq 3$ moduli space orbifold geometries at rank 2}},
  \href{http://arxiv.org/abs/1904.10969}{{\tt arXiv:1904.10969}}.

\bibitem{Argyres:2019yyb}
P.~C. Argyres, A.~Bourget, and M.~Martone, {\it {On the moduli spaces of 4d
  $\mathcal{N} = 3$ SCFTs I: triple special K{\"a}hler structure}},
  \href{http://arxiv.org/abs/1912.04926}{{\tt arXiv:1912.04926}}.

\bibitem{Aharony:2013hda}
O.~Aharony, N.~Seiberg, and Y.~Tachikawa, {\it {Reading between the lines of
  four-dimensional gauge theories}},  {\em JHEP} {\bf 08} (2013) 115,
  [\href{http://arxiv.org/abs/1305.0318}{{\tt arXiv:1305.0318}}].

\bibitem{Bourget:2018ond}
A.~Bourget, A.~Pini, and D.~Rodr\'iguez-G\'omez, {\it {The Importance of Being
  Disconnected, A Principal Extension for Serious Groups}},
  \href{http://arxiv.org/abs/1804.01108}{{\tt arXiv:1804.01108}}.

\bibitem{Argyres:2018zay}
P.~C. Argyres, C.~Long, and M.~Martone, {\it {The Singularity Structure of
  Scale-Invariant Rank-2 Coulomb Branches}},  {\em JHEP} {\bf 05} (2018) 086,
  [\href{http://arxiv.org/abs/1801.01122}{{\tt arXiv:1801.01122}}].

\bibitem{Seiberg:1994rs}
N.~Seiberg and E.~Witten, {\it {Electric - magnetic duality, monopole
  condensation, and confinement in N=2 supersymmetric Yang-Mills theory}},
  {\em Nucl. Phys.} {\bf B426} (1994) 19--52,
  [\href{http://arxiv.org/abs/hep-th/9407087}{{\tt hep-th/9407087}}]. [Erratum:
  Nucl. Phys.B430,485(1994)].

\bibitem{Seiberg:1994aj}
N.~Seiberg and E.~Witten, {\it {Monopoles, duality and chiral symmetry breaking
  in N=2 supersymmetric QCD}},  {\em Nucl. Phys.} {\bf B431} (1994) 484--550,
  [\href{http://arxiv.org/abs/hep-th/9408099}{{\tt hep-th/9408099}}].

\bibitem{AlvarezGaume:1996mv}
L.~Alvarez-Gaume and S.~F. Hassan, {\it {Introduction to S duality in N=2
  supersymmetric gauge theories: A Pedagogical review of the work of Seiberg
  and Witten}},  {\em Fortsch. Phys.} {\bf 45} (1997) 159--236,
  [\href{http://arxiv.org/abs/hep-th/9701069}{{\tt hep-th/9701069}}].

\bibitem{Lerche:1996xu}
W.~Lerche, {\it {Introduction to Seiberg-Witten theory and its stringy
  origin}},  {\em Nucl. Phys. Proc. Suppl.} {\bf 55B} (1997) 83--117,
  [\href{http://arxiv.org/abs/hep-th/9611190}{{\tt hep-th/9611190}}]. [Nucl.
  Phys. Proc. Suppl.55,83(1997)].

\bibitem{Argyres:1996nonp}
P.~C. Argyres, ``{\emph{Non-Perturbative Dynamics of Four-Dimensional
  Supersymmetric Field Theories}}.''
  \url{https://homepages.uc.edu/~argyrepc/cu661-gr-SUSY/fgilec.pdf}, 1998.

\bibitem{Ramond:1999susy}
P.~Ramond, {\em Journeys Beyond the Standard Model}.
\newblock Frontiers in Physics, 1999.

\bibitem{Weinberg:2000III}
S.~Weinberg, {\em The Quantum Theory of Fields, Vol. 3: Supersymmetry}.
\newblock Cambridge University Press, 2000.

\bibitem{Terning:2006ms}
J.~Terning, {\em Modern Supersymmetry: Dynamics and Duality}.
\newblock Oxford University Press, 2006.

\bibitem{Dine:2007Sst}
M.~Dine, {\em Supersymmetry and String Theory: Beyond the Standard Model}.
\newblock Cambridge University Press, 2007.

\bibitem{Shifman:2012lec}
M.~A. Shifman, {\em Advanced Topics in quantum field theory: a lecture course}.
\newblock Cambridge University Press, 2007.

\bibitem{Gates:1983nr}
S.~J. Gates, M.~T. Grisaru, M.~Rocek, and W.~Siegel, {\it {Superspace Or One
  Thousand and One Lessons in Supersymmetry}},  {\em Front. Phys.} {\bf 58}
  (1983) 1--548, [\href{http://arxiv.org/abs/hep-th/0108200}{{\tt
  hep-th/0108200}}].

\bibitem{Shifman:1995ua}
M.~A. Shifman, {\it {Nonperturbative dynamics in supersymmetric gauge
  theories}},  {\em Prog. Part. Nucl. Phys.} {\bf 39} (1997) 1--116,
  [\href{http://arxiv.org/abs/hep-th/9704114}{{\tt hep-th/9704114}}].
  [,345(1995)].

\bibitem{Martin:1997ns}
S.~P. Martin, {\it {A Supersymmetry primer}},
  \href{http://arxiv.org/abs/hep-ph/9709356}{{\tt hep-ph/9709356}}. [Adv. Ser.
  Direct. High Energy Phys.18,1(1998)].

\bibitem{Argyres:2001eva}
P.~C. Argyres, ``{\emph{An Introduction to Global Supersymmetry}}.''
  \url{https://homepages.uc.edu/~argyrepc/cu661-gr-SUSY/susy2001.pdf}, 2001.

\bibitem{Luty:2005sn}
M.~A. Luty, {\it {2004 TASI lectures on supersymmetry breaking}},  in {\em
  {Physics in D >= 4. Proceedings, Theoretical Advanced Study Institute in
  elementary particle physics, TASI 2004, Boulder, USA, June 6-July 2, 2004}},
  pp.~495--582, 2005.
\newblock \href{http://arxiv.org/abs/hep-th/0509029}{{\tt hep-th/0509029}}.

\bibitem{Aitchison:2005cf}
I.~J.~R. Aitchison, {\it {Supersymmetry and the MSSM: An Elementary
  introduction}},  \href{http://arxiv.org/abs/hep-ph/0505105}{{\tt
  hep-ph/0505105}}.

\bibitem{Tachikawa:2013kta}
Y.~Tachikawa, {\em {N=2 supersymmetric dynamics for pedestrians}}, vol.~890.
\newblock 2014.

\bibitem{Argyres:1996eh}
P.~C. Argyres, M.~R. Plesser, and N.~Seiberg, {\it {The Moduli space of vacua
  of N=2 SUSY QCD and duality in N=1 SUSY QCD}},  {\em Nucl. Phys.} {\bf B471}
  (1996) 159--194, [\href{http://arxiv.org/abs/hep-th/9603042}{{\tt
  hep-th/9603042}}].

\bibitem{Gaiotto:2008nz}
D.~Gaiotto, A.~Neitzke, and Y.~Tachikawa, {\it {Argyres-Seiberg duality and the
  Higgs branch}},  {\em Commun. Math. Phys.} {\bf 294} (2010) 389--410,
  [\href{http://arxiv.org/abs/0810.4541}{{\tt arXiv:0810.4541}}].

\bibitem{mckay1981tables}
W.~McKay and J.~Patera, {\em {Tables of dimensions, indices, and branching
  rules for representations of simple Lie algebras}}.
\newblock Lecture notes in pure and applied mathematics, M. Dekker, 1981.

\bibitem{Seiberg:1994pq}
N.~Seiberg, {\it {Electric - magnetic duality in supersymmetric nonAbelian
  gauge theories}},  {\em Nucl. Phys.} {\bf B435} (1995) 129--146,
  [\href{http://arxiv.org/abs/hep-th/9411149}{{\tt hep-th/9411149}}].

\bibitem{Argyres:2017tmj}
P.~C. Argyres, Y.~L{\"u}, and M.~Martone, {\it {Seiberg-Witten geometries for
  Coulomb branch chiral rings which are not freely generated}},  {\em JHEP}
  {\bf 06} (2017) 144, [\href{http://arxiv.org/abs/1704.05110}{{\tt
  arXiv:1704.05110}}].

\bibitem{Argyres:2018wxu}
P.~C. Argyres and M.~Martone, {\it {Coulomb branches with complex
  singularities}},  {\em JHEP} {\bf 06} (2018) 045,
  [\href{http://arxiv.org/abs/1804.03152}{{\tt arXiv:1804.03152}}].

\bibitem{Montonen:1977sn}
C.~Montonen and D.~I. Olive, {\it {Magnetic Monopoles as Gauge Particles?}},
  {\em Phys. Lett. B} {\bf 72} (1977) 117--120.

\bibitem{Cardy:1981qy}
J.~L. Cardy and E.~Rabinovici, {\it {Phase Structure of Z(p) Models in the
  Presence of a Theta Parameter}},  {\em Nucl. Phys. B} {\bf 205} (1982) 1--16.

\bibitem{Cardy:1981fd}
J.~L. Cardy, {\it {Duality and the Theta Parameter in Abelian Lattice Models}},
   {\em Nucl. Phys. B} {\bf 205} (1982) 17--26.

\bibitem{Shapere:1988zv}
A.~D. Shapere and F.~Wilczek, {\it {Selfdual Models with Theta Terms}},  {\em
  Nucl. Phys. B} {\bf 320} (1989) 669--695.

\bibitem{Freed:1997dp}
D.~S. Freed, {\it {Special Kahler manifolds}},  {\em Commun. Math. Phys.} {\bf
  203} (1999) 31--52, [\href{http://arxiv.org/abs/hep-th/9712042}{{\tt
  hep-th/9712042}}].

\bibitem{Cordova:2016emh}
C.~Cordova, T.~T. Dumitrescu, and K.~Intriligator, {\it {Multiplets of
  Superconformal Symmetry in Diverse Dimensions}},
  \href{http://arxiv.org/abs/1612.00809}{{\tt arXiv:1612.00809}}.

\bibitem{Minahan:1996fg}
J.~A. Minahan and D.~Nemeschansky, {\it {An N=2 superconformal fixed point with
  E(6) global symmetry}},  {\em Nucl.Phys.} {\bf B482} (1996) 142--152,
  [\href{http://arxiv.org/abs/hep-th/9608047}{{\tt hep-th/9608047}}].

\bibitem{Minahan:1996cj}
J.~A. Minahan and D.~Nemeschansky, {\it {Superconformal fixed points with E(n)
  global symmetry}},  {\em Nucl.Phys.} {\bf B489} (1997) 24--46,
  [\href{http://arxiv.org/abs/hep-th/9610076}{{\tt hep-th/9610076}}].

\bibitem{Xie:2012hs}
D.~Xie, {\it {General Argyres-Douglas Theory}},  {\em JHEP} {\bf 01} (2013)
  100, [\href{http://arxiv.org/abs/1204.2270}{{\tt arXiv:1204.2270}}].

\bibitem{Argyres:2018urp}
P.~C. Argyres and M.~Martone, {\it {Scaling dimensions of Coulomb branch
  operators of 4d N=2 superconformal field theories}},
  \href{http://arxiv.org/abs/1801.06554}{{\tt arXiv:1801.06554}}.

\bibitem{Caorsi:2018zsq}
M.~Caorsi and S.~Cecotti, {\it {Geometric classification of 4d $\mathcal{N}=2$
  SCFTs}},  {\em JHEP} {\bf 07} (2018) 138,
  [\href{http://arxiv.org/abs/1801.04542}{{\tt arXiv:1801.04542}}].

\bibitem{Griff:1978}
P.~Griffiths and J.~Harris, {\em Principles of Algebraic Geometry}.
\newblock Wiley, 1978.

\bibitem{Shafa:1977}
I.~R. Shafarevich, {\em Basic Algebraic Geomety 1 $\&$ 2}.
\newblock Springer, 1977.

\bibitem{Argyres:2005pp}
P.~C. Argyres, M.~Crescimanno, A.~D. Shapere, and J.~R. Wittig, {\it
  {Classification of N=2 superconformal field theories with two-dimensional
  Coulomb branches}},  \href{http://arxiv.org/abs/hep-th/0504070}{{\tt
  hep-th/0504070}}.

\bibitem{Argyres:2005wx}
P.~C. Argyres and J.~R. Wittig, {\it {Classification of N=2 superconformal
  field theories with two-dimensional Coulomb branches. II.}},
  \href{http://arxiv.org/abs/hep-th/0510226}{{\tt hep-th/0510226}}.

\bibitem{Klemm:1994qs}
A.~Klemm, W.~Lerche, S.~Yankielowicz, and S.~Theisen, {\it {Simple
  singularities and N=2 supersymmetric Yang-Mills theory}},  {\em Phys. Lett.}
  {\bf B344} (1995) 169--175, [\href{http://arxiv.org/abs/hep-th/9411048}{{\tt
  hep-th/9411048}}].

\bibitem{Argyres:1994xh}
P.~C. Argyres and A.~E. Faraggi, {\it {The vacuum structure and spectrum of N=2
  supersymmetric SU(n) gauge theory}},  {\em Phys. Rev. Lett.} {\bf 74} (1995)
  3931--3934, [\href{http://arxiv.org/abs/hep-th/9411057}{{\tt
  hep-th/9411057}}].

\bibitem{Argyres:1995fw}
P.~C. Argyres and A.~D. Shapere, {\it {The Vacuum structure of N=2 superQCD
  with classical gauge groups}},  {\em Nucl. Phys.} {\bf B461} (1996) 437--459,
  [\href{http://arxiv.org/abs/hep-th/9509175}{{\tt hep-th/9509175}}].

\bibitem{Argyres:1995wt}
P.~C. Argyres, M.~R. Plesser, and A.~D. Shapere, {\it {The Coulomb phase of N=2
  supersymmetric QCD}},  {\em Phys. Rev. Lett.} {\bf 75} (1995) 1699--1702,
  [\href{http://arxiv.org/abs/hep-th/9505100}{{\tt hep-th/9505100}}].

\bibitem{Danielsson:1995is}
U.~H. Danielsson and B.~Sundborg, {\it {The Moduli space and monodromies of N=2
  supersymmetric SO(2r+1) Yang-Mills theory}},  {\em Phys. Lett.} {\bf B358}
  (1995) 273--280, [\href{http://arxiv.org/abs/hep-th/9504102}{{\tt
  hep-th/9504102}}].

\bibitem{Hanany:1995na}
A.~Hanany and Y.~Oz, {\it {On the quantum moduli space of vacua of N=2
  supersymmetric SU(N(c)) gauge theories}},  {\em Nucl. Phys.} {\bf B452}
  (1995) 283--312, [\href{http://arxiv.org/abs/hep-th/9505075}{{\tt
  hep-th/9505075}}].

\bibitem{Brandhuber:1995zp}
A.~Brandhuber and K.~Landsteiner, {\it {On the monodromies of N=2
  supersymmetric Yang-Mills theory with gauge group SO(2n)}},  {\em Phys.
  Lett.} {\bf B358} (1995) 73--80,
  [\href{http://arxiv.org/abs/hep-th/9507008}{{\tt hep-th/9507008}}].

\bibitem{Donagi:1995cf}
R.~Donagi and E.~Witten, {\it {Supersymmetric Yang-Mills theory and integrable
  systems}},  {\em Nucl. Phys.} {\bf B460} (1996) 299--334,
  [\href{http://arxiv.org/abs/hep-th/9510101}{{\tt hep-th/9510101}}].

\bibitem{Witten:1988ze}
E.~Witten, {\it {Topological Quantum Field Theory}},  {\em Commun. Math. Phys.}
  {\bf 117} (1988) 353.

\bibitem{Hitchin:1987mz}
N.~J. Hitchin, {\it {Stable bundles and integrable systems}},  {\em Duke Math.
  J.} {\bf 54} (1987) 91--114.

\end{thebibliography}

\end{document}